\documentclass[11pt]{article}

\usepackage{authblk}
\author{Daniil Bargman\thanks{daniil.bargman.22@ucl.ac.uk}}
\author{Francesca Medda\thanks{f.medda@ucl.ac.uk}}
\author{Akash Sedai Sharma\thanks{akash.sharma@ucl.ac.uk}}
\affil{Institute of Finance and Technology, University College London}
\usepackage{geometry}
\usepackage{natbib}
\setcitestyle{authoryear,round}

\let\citep\citet
\usepackage[breaklinks]{hyperref}
\hypersetup{colorlinks=true,urlcolor=blue,linkcolor=blue,citecolor=blue}
\usepackage{multirow}
\usepackage{adjustbox}
\usepackage{threeparttable}
\usepackage{caption}
\usepackage{booktabs}
\usepackage{float}
\usepackage{pdflscape}
\usepackage{bm}
\usepackage{amsfonts}
\usepackage{amsthm}
\usepackage{nccmath}
\usepackage{cases}
\usepackage{stackengine}
\usepackage{geometry}
\usepackage{mathtools}
\usepackage{caption}
\usepackage{placeins}
\usepackage{threeparttable}

\let\citep\citet
\newcommand{\brac}[1]{\left( {#1} \right)}
\def\y{\mathbf{\bm{y}}}

\def\P{\mathbf{\bm{P}}}
\def\B{\mathbf{\bm{\beta}}}

\def\tx{\tilde{x}}

\def\1{\mathbf{\bm{1}}}
\def\0{\mathbf{\bm{0}}}

\def\Sm{\mathbf{\bm{\Sigma}}}

\def\o{\bm{\omega}}
\def\bo{\brac{\B \otimes \o}}
\def\bi{\brac{\B \otimes I_{n}}}
\def\bio{\bi \o}
\def\oi{\brac{I_{F} \otimes \o}}
\def\oib{\oi \B}

\setstackgap{L}{8pt}

\newcommand{\E}[2][]{\ensuremath{\mathbb{E}{#1} \left[ {#2} \right]}}
\usepackage{enumitem}
\renewlist{itemize}{itemize}{9}
\setlist[itemize]{label=$\circ$}
\date{\today}
\title{Latent Variable Phillips Curve}
\hypersetup{
 pdfauthor={},
 pdftitle={Latent Variable Phillips Curve},
 pdfkeywords={},
 pdfsubject={},
 pdfcreator={},
 pdflang={English}}
\begin{document}

\maketitle
\begin{abstract}

This paper re-examines the empirical Phillips curve (PC) model and its
usefulness in the context of medium-term inflation forecasting. A latent
variable Phillips curve hypothesis is formulated and tested using 3,968
randomly generated factor combinations. Evidence from US core PCE
inflation between Q1 1983 and Q1 2025 suggests that latent variable PC
models reliably outperform traditional PC models six to eight quarters
ahead and stand a greater chance of outperforming a univariate
benchmark. Incorporating an MA(1) residual process improves the accuracy
of empirical PC models across the board, although the gains relative to
univariate models remain small. The findings presented in this paper
have two important implications: First, they corroborate a new
conceptual view on the Phillips curve theory; second, they offer a novel
path towards improving the competitiveness of Phillips curve forecasts
in future empirical work.

\end{abstract}
\section{Introduction}
\label{introduction}
Medium-term inflation targeting lies at the heart of modern monetary
policy. A widely accepted interpretation of the words ``medium term'' is
a time horizon of around two years (e.g., see \citep{Hammond-2012}),
which means that inflation forecasts for monetary policy decisions need
to cover 8 quarterly observations, or 24 observations in case of monthly
projections.

A large body of academic literature examines various methodologies for
forecasting inflation. One of the most popular models is the Phillips
curve (PC) introduced by \citep{Phillips-1958} and popularised by
\citep{Samuelson-Solow-1960}. Classical Phillips curve models link
future inflation to past and present economic conditions approximated by
aggregates such as labour market slack or the output gap. Many PC models
also include inflation expectations (e.g., \citep{Ball-1988,Hommes-1998,Lansing-2009,Albuquerque-2017}), supply shocks (e.g.,
\citep{Gordon-1977a,Chen-2014,Salisu-2018}), or other control
variables such as financial conditions (e.g., \citep{Conti-2021}).

Despite their popularity among economic theorists, Phillips curve models
have historically shown mixed results as a tool for empirical inflation
forecasting. Following a great effort over the past 70 years to extend
the Phillips curve theory into the edge cases such as wars, commodity
market disruptions and other unstable economic environments (see
\citep{Gordon-2011} for a historical overview), most researchers still
find that the Phillips curve either changes shape or breaks down
entirely under certain conditions (\citep{Conti-2021,Mutascu-2019,Albuquerque-2017,Ball-2011,Stock-2007,Stock-1999}), and many find
that PC forecasts of inflation offer little to no improvement over
ARIMA-style univariate techniques (e.g., \citep{Faust-2013,Ang-2007}).

It is likely that at least some of the reasons for the lacklustre
empirical performance of Phillips curve models are unrelated to the
merits of the underlying theory itself. Here, at least three
possibilities come to mind. First, as with all macroeconomic aggregates,
the average accuracy of inflation forecasts will inevitably decline over
longer horizons due to an increased likelihood of unforeseen economic
shocks. Second, policymakers and economists have long speculated that
``long and variable lags'' exist between changes in monetary policy
(and, by extension, changes in economic conditions) and changes in
inflation\footnote{This idea can be traced back to \cite{Friedman-1960-book} but has
since resonated in a number of speeches given by various monetary policy
officials, including Fed Chair Powell's press conference on 2
November 2022.}. Third, historical economic data is often highly
imprecise to begin with: measurement errors, biased sampling techniques,
seasonal adjustment issues, and methodology revisions\footnote{Two relevant examples of significant changes in calculation
methodology include a switch to owner-equivalent rents as a proxy of US
housing inflation in 1983, and the change to the Employment Cost Index
(ECI) methodology in 2006.} are just a
few examples. It is highly likely that these empirical limitations
explain at least some of the real-world challenges faced by the Phillips
curve and other parameter-rich inflation models\footnote{For example, \cite{Faust-2013} performs a large-scale survey of
inflation forecasting techniques and concludes that the best-performing
forecasts of inflation are based on surveys rather than statistical
models. However, even survey-based forecasts are only found to improve
on the root mean squared prediction error (RMSPE) of the naïve forecast
by less than 20\% zero to four quarters ahead. In terms of this paper's
chosen metric of forecasting accuracy (MSPE, i.e., the square of RMSPE),
this would translate into an improvement of less than 10\%.}, as well as why
these models struggle against the simplest univariate forecasting
approaches. Put simply, the more complex models are more likely to
over-fit the data when the signal-to-noise ratio is poor.

Several prior studies have recognised this problem and attempted to
mitigate it, with some success, using a combination of noise reduction
and model reduction techniques. For example, \citep{Conti-2021} is able
to estimate a more stable Phillips curve in Italy by constructing custom
composite measures of labour market slack and financial conditions;
\citep{Stock-1999} uses information criteria to determine the optimal
lag structure in empirical PC models, as well as constructing composite
explanatory variables from individual real activity measures;
\citep{Berriel-2016} considers several feature selection techniques for
Phillips curve models, advocating for LASSO as the most effective
methodology.

This paper takes empirical PC research further in the direction of both
model parsimony and synthetic explanatory variables. The Phillips curve
is re-interpreted as an inherently univariate relationship, with the
caveat that the ``true'' driving force behind inflation is unobserved
(latent). Specifically, it is hypothesised that the observed explanatory
variables put forward by traditional Phillips curve theories are simply
imperfect approximations of a singular unobserved process, loosely
interpreted as the aggregate amount of price pressure in the real
economy. This hypothesis is transformed into a tractable empirical model
using a novel regression methodology called Latent Shock Regression
(LSR), introduced in \citep{Bargman-clarx-2025-preprint}.

The primary goal of this paper is to test the latent variable Phillips
curve (LVPC) hypothesis. The empirical section examines whether a
randomly selected Phillips curve factor combination would generate
superior medium-term inflation projections using the LSR methodology.
3,968 randomly generated PC factor combinations are used in an
out-of-sample horse race between LSR-based LVPC models, traditional PC
models, and a standard set of univariate benchmarks. The comparison is
made without any real-time model optimisation or feature selection, in
order to obtain transparent like-for-like model performance statistics.
In other words, this paper makes no explicit attempt to derive the
single best empirical model of inflation, but rather opts to re-assess
the Phillips curve from a conceptual perspective and present a path
towards improving its competitiveness in real-world settings.

Initial findings speak in favour of the latent variable Phillips curve
hypothesis: LSR-based LVPC models overwhelmingly outperform traditional
PC models in medium-term inflation forecasting. The gains in accuracy
compared to the univariate benchmarks remain statistically and
economically small in aggregate, which is both consistent with prior
research and somewhat expected in the absence of model calibration
overlays.

The rest of this paper is organised as follows. Section
\ref{literature_review} provides an overview of the relevant empirical
Phillips curve literature and contextualises the contribution of this
paper. Section \ref{methodology} formulates the latent variable Phillips
curve hypothesis and describes the LSR methodology. In Section
\ref{empirical_results}, the LVPC hypothesis is put to a series of
statistical tests. Section \ref{concluding_remarks} concludes and translates
the main empirical findings into a list of tangible methodological
suggestions for improving the competitiveness of Phillips curve models
in future empirical work.
\section{Background}
\label{literature_review}
The Phillips curve (PC) gets its name from \citep{Phillips-1958} which
discovers an inverse relationship between unemployment and nominal wage
growth in the UK between 1861 and 1957. Phillips argued that inflation
stems from the difference between nominal wage growth and productivity
growth. The idea was subsequently picked up by scholars in the U.S.
(notably, \citep{Samuelson-Solow-1960}), resulting in two alternative
schools of thought and two competing variants of the PC model: the
``triangle'' model with sticky inflation and commodity supply shocks,
and the new Keynesian model (NKPC) in which inflation is additionally
affected by inflation expectations\footnote{See \cite{Gordon-2011} for an overview of the history of the
Phillips curve theory.}.

A large body of empirical literature makes a pragmatic attempt to use
the Phillips curve to forecast inflation. Rather than inferring
inflation from wages, empirical PC models usually attempt to forecast
inflation directly. One of the most widely cited papers in the field is
\citep{Stock-1999} whose contribution is largely threefold: Firstly, it
provides one of the first comprehensive surveys over a large number of
explanatory variables of inflation, examining nearly 200 regression
candidates for three alternative measures of inflation in the United
States. As such, an argument is made in favour of a more flexible
interpretation of the Phillips curve in which unemployment is merely a
proxy for the broader concept of real economic activity. Secondly, the
paper attempts to construct purpose-built synthetic measures of activity
by combining multiple economic indicators into a single explanatory
variable. A composite indicator comprised of a combination of real
activity measures is ultimately found to be the best predictor of
inflation out of sample. Third, the paper identifies a structural break
in the predictability of inflation in 1983 resulting from a change in
the measurement of owner-occupied housing costs.

Although many of the findings from \citep{Stock-1999} have generally
stood the test of time, the evidence favouring empirical PC models over
the other inflation forecasting techniques has since been called into
question. For example, \citep{Ang-2007} performs a comprehensive
comparison of inflation forecasting techniques including Phillips curve
models, univariate time series (ARIMA) models, factor models based on
the term structure of interest rates, and factor models using
survey-based inflation forecasts (SPF, Livingston, and Michigan), and
finds that survey-based forecasts are the best predictor of inflation,
whereas Phillips curve models do not reliably improve upon an ARMA(1,1)
process or even a simple random walk. \citep{Faust-2013} performs
another comprehensive comparison of 17 broad types of inflation
forecasting models including the Phillips curve, a vector autoregressive
model based on the term structure of interest rates, dynamic structural
general equilibrium models, and a Bayesian model average over 77
different alternative predictors, concluding once again that qualitative
forecasts of inflation (the Blue Chip survey, the Survey of Professional
Forecasters and the Fed staff's Greenbook forecast) are the only
forecasts that reliably outperform a univariate benchmark.

This new evidence seems to have led to a type of bifurcation in the
empirical PC literature, with one body of papers attempting to identify
the root causes of the mixed performance of PC models, and another
attempting to improve on the accuracy of PC forecasts using various
statistical overlays. In the former camp, several empirical papers point
to a lack of stability in the Phillips curve relationship. For example,
\citep{Peach-2011} argues that economic slack only influences inflation
beyond certain thresholds. \citep{Albuquerque-2017} argues that the
slope of the Phillips curve is time-varying, which may cause inflation
to rise suddenly and by more than expected -- a proposition that has
aged well in light of the global inflation spike in 2022.
\citep{Mutascu-2019} studies the Phillips Curve using wavelets and
concludes that the relationship breaks down or even reverses sign during
periods of economic stress (recessions, oil price shocks, stagflations)
and strong monetary policy interventions. The theoretical justifications
for these results are mostly rooted in bounded rationality and the
resulting bias in inflation expectations, e.g., \citep{Ball-2000},
\citep{Lansing-2009} and \citep{Casarin-2025} identify regimes in
inflation expectations corresponding to different periods in US economic
history.

In the latter camp, \citep{Conti-2021} is able to improve the accuracy
of inflation forecasts in Italy by using a composite indicator of labour
market slack, similar to what \citep{Stock-1999} does with real economic
activity in the US, and by introducing a composite measure of
market-based financial conditions as a control. In contrast to
\citep{Albuquerque-2017}, \citep{Conti-2021} finds that a ``better''
proxy for economic slack is able to produce a more stable Phillips
curve. \citep{Chen-2014} and \citep{Salisu-2018} look at commodity
prices as predictors of inflation, and both find evidence of their
relevance, contrary to the initial finding in \citep{Stock-1999}.
\citep{Berriel-2016} finds that the empirical fit of Phillips curve
models can be improved with a LASSO-based instrument selection overlay.
\citep{Lansing-2019} reports an improvement in the predictive power of a
standard PC model after introducing an interaction term between lags of
inflation and the output gap.

This paper seeks to contribute to this latter camp of empirical
research. The Phillips curve theory is looked at from a new angle by
proposing that the ``true'' economic driver of inflation is an
unobserved aggregate price pressure process in the real economy. Similar
to \citep{Stock-1999} and \citep{Conti-2021}, various combinations of
explanatory variables are used to ``synthesise'' alternative proxies for
this price pressure process. It is shown that these synthetic proxies
generate more accurate medium-term forecasts of inflation in the vast
majority of cases.
\section{Methodology}
\label{methodology}
Inflation represents the sumtotal of price-setting decisions made by the
producers of goods and services in an economy. Each individual
price-setting decision is influenced by agent- (\citep{Sims-2006}) and
industry-specific considerations (\citep{Bryan-2010}), as well as a
common macroeconomic force. The common force can be loosely defined as
the aggregate price pressure in the real economy, although it has been
described by several other names in prior Phillips curve
literature\footnote{Many Phillips curve papers use the generic term ``slack'',
referring loosely to gaps in the labour market or capacity utilisation;
\cite{Ball-1988} uses the term ``fluctuations in aggregate demand'';
\cite{Lansing-2009} uses a non-descript ``stationary driving variable''.}. Either way, it represents a confluence of
socio-economic circumstances leading large groups of producers to change
prices in the same direction, albeit in a staggered fashion owing to the
associated costs\footnote{This seems to be the accepted explanation within the new
Keynesian Phillips curve framework -- see \citep{Ball-1988}.}.

The ``true'' aggregate price pressure process in the real economy is
an inherently unobserved phenomenon. However, an argument can be made
that the observed variables that lead inflation in traditional Phillips
curve models (e.g., the reported gaps in unemployment and output,
surveys of inflation expectations, changes in commodity prices, even
past inflation itself) are either the underlying drivers of this
unobserved process, or merely its timelier approximations.

This idea can be expressed formally as a latent variable Phillips curve
hypothesis. First, let us represent the unobserved aggregate price
pressure process in the real economy as a latent variable \(\tx\) and
assume that it leads inflation -- an observed dependent variable \(y\)
-- up to \(F\) periods in the future. The ``traditional'' predictors of
inflation proposed by the Phillips curve theory, which we place in a
vector \(X = \begin{bmatrix} x_1, & x_2, & x_3, & \ldots, & x_n
\end{bmatrix}\), each provide some relevant information about \(\tx\).
The latent variable Phillips Curve hypothesis would then state that
there exists a vector of weights \(\o\) with dimensions \(n \times 1\)
such that the linear combination \(X \o\) approximates \(\tx\), and
hence predicts \(y\) up to \(F\) periods ahead, better than the
individual elements of \(X\).

To test this hypothesis empirically, define the following regression
problem:

\begin{equation}\label{eqn:lsr}
\begin{split}
& y_t = c + \sum_{\tau = 1}^{F} \beta_{\tau} \tx_{t-\tau} + \epsilon_t \\
& \tx_t = X_t \o + r_t \text{ for all } t
\end{split}
\end{equation}

Here, \(t \in \mathbb{N}\) denotes the realization of a random variable
at time \(t\), \(F \in \mathbb{N}\) is the maximum lag order of \(\tx\)
(and, conversely, the maximum forecast horizon over which \(\tx\) is
assumed to affect \(y\)), and \(\epsilon\) and \(r\) are error terms.
The same regression model can be expressed as a single equation in
matrix form:

\begin{equation}\label{eqn:lsr_vec}
y_t = c + P \left( \B \otimes \o \right) + \epsilon_t
\end{equation}

where \(P \equiv \begin{bmatrix}L^1X, & L^2X, & L^3X, & \ldots, & L^FX
\end{bmatrix}\) is a \(1 \times nF\) vector containing all the lags of
\(X\), \(\B = \begin{bmatrix}\beta_1, & \beta_2, & \beta_3, & \ldots, &
\beta_F \end{bmatrix}'\) is an \(F \times 1\) column vector containing
all the time series coefficients \(\beta_{\tau}\), and \(\otimes\) is
the Kronecker product.

The mathematics of this regression model is examined in more detail in
\citep{Bargman-clarx-2025-preprint} under the name ``Latent Shock
Regression'' (LSR), which is a special case of a more complex regression
methodology called (C)LARX -- (constrained) latent variable
autoregression with exogenous inputs. For completeness, the solution for
LSR is outlined in \ref{lsr_derivation}. It is given by:

\begin{equation}\label{eqn:lsr_solution}
\begin{split}
& \B = \left[\oi' \Sm_{P} \oi \right]^{-1} \oi' \Sm_{Py} \\
& \o = \left[\bi' \Sm_{P} \bi \right]^{-1} \bi' \Sm_{Py} \\
& c = \overline{\y} - \overline{\P} \bo
\end{split}
\end{equation}

Here, \(I_{n}\) and \(I_F\) denote identity matrices of size \(n\) and
\(F\), respectively, \(\Sm_{P}\) and \(\Sm_{Py}\) denote the sample
covariance matrix of \(\P\) and the covariance from \(\P\) to \(\y\),
\(\overline{\y}\) is the sample mean of \(y\), and \(\overline{\P}\) is
a \(1 \times nF\) row vector containing the column-wise sample means in
\(\P\).

This formula may look a little involved at first, but it simply
expresses \(\B\) as the OLS solution for (\ref{eqn:lsr}) conditional on the
value of \(\o\), and \(\o\) as the OLS solution to (\ref{eqn:lsr}) conditional
on the value of \(\B\). This full system can be solved using standard
fixed point iteration\footnote{First, make an initial guess for one of the vectors, e.g., guess
\(\o_0\) to be the value of vector \(\o\). Plug \(\o_0\) into the
formula for \(\B\), obtaining an estimate \(\B_1\). Then plug \(\B_1\)
back into the formula for \(\o\), obtaining an updated guess \(\o_1\)
for \(\o\). Plug \(\o_1\) back into the formula for \(\B\), updating the
estimate for \(\B\) to \(\B_2\). Repeat the process until \(\o_{k+1}\)
and \(\B_{k+1}\) are identical to \(\o_k\) and \(\B_k\), respectively,
within a predefined tolerance level.} as long as both \(\left(I_F \otimes \o \right)'
\Sm_{P} \left(I_F \otimes \o \right)\) and \(\left(\B \otimes I_n
\right)' \Sm_{P} \left(\B \otimes I_n \right)\) remain invertible at
each iteration step. An initial guess is required for either \(\o\) or
\(\B\), but it can be as simple as a unit vector of the appropriate
length (\(F\) for \(\B\), \(n\) for \(\o\)) because the formula is
scale-invariant with respect to either \(\o\) or \(\B\) by the
properties of the Kronecker product.

The LSR methodology represents a tractable empirical implementation of a
latent variable Phillips curve model, but it also has some noteworthy
statistical properties in its own right. Like traditional lead-lag
regressions, LSR is a least-squares multiple regression between a
dependent variable, \(y\), and lags of a vector of explanatory variables
\(X\). In fact, it follows directly from (\ref{eqn:lsr_solution}) that LSR
reduces to a traditional lead-lag regression when either \(n\) or \(F\)
is 1. However, because LSR treats all \(x_i\) in \(X\) as facets of a
single information process, an implicit assumption is imposed that all
\(x_i\) affect \(y\) with the same lag profile. This creates a
separation of concerns between the cross-sectional dimension of
regression responses, represented by a time-invariant \(\o\), and the
time series dimension of responses, represented by \(\B\). The
coefficient for variable \(x_i\) at lag \(\tau\) is then given by the
product of the \(i\)'th element of \(\o\) and the \(\tau\)'th element of
\(\B\).

While this structural constraint may lead to a poorer in-sample fit for
LSR models compared to traditional lead-lag regressions, it offers a
considerable benefit of parsimony in multivariate, multi-periods
forecasting settings. To illustrate, consider a standard lead-lag
regression of \(y\) on a series of lagged explanatory variables \(x_i\)
used in traditional empirical Phillips curve models:

\begin{equation}\label{eqn:lead_lag_multi}
y_t = c + \sum_{i = 1}^{n}\sum_{\tau = 1}^{F} \beta_{i,\tau}x_{i,t-\tau} +\epsilon_t
\end{equation}

The number of coefficients that need to be estimated with this
methodology is \(1 + n \times F\) in the upper bound\footnote{The term ``upper bound'' is an homage to feature selection.}, where \(n\)
is the number of explanatory variables and \(F\) is the forecast
horizon. Alternatively, a direct forecast could be attempted for each
forecast horizon by running \(F\) separate regressions, such that:

\begin{equation}\label{eqn:lead_lag_multi_direct}
y_{t+\tau} = c + \sum_{i = 1}^{n}\beta_{i,\tau}x_{i,t} +\epsilon_t \quad
\text{for } \tau \in \left[ 1, F \right]
\end{equation}

This reduces the number of regression coefficients in a single model to
\(1 + n\), but the total number of coefficients required for a full
forecast profile is now \(F \left( 1 + n  \right) = F + n \times
F\).

In contrast, an LSR model with the same number of observed explanatory
variables \(n\) and forecast horizon \(F\) will only require an \(n
\times 1\) vector of cross-sectional responses \(\o\), an \(F \times 1\)
vector of time series responses, \(\B\), and an intercept term \(c\).
This results in \(1 + n + F\) coefficients compared to \(1 + n \times
F\) coefficients in a traditional specification. The difference may seem
subtle at first, but it does gain in importance when both \(n\) and
\(F\) are above 1 and at least one of them becomes large. For example,
let us assume that an economist at a central bank wishes to generate
inflation projections up to 8 quarters ahead (\(F = 8\)) using the
standard ``triangle'' Phillips curve model (\(n = 3\)). In this case,
the upper-bound number of coefficients required by a traditional
lead-lag regression methodology would be 25 (or 32 with a direct
forecasting approach), whereas an LSR regression would only require 12.
\subsection{Note: Out-of-Sample Predictions with LSR}
\label{sec:org3eda785}

A real-life LSR forecast for the dependent variable at time \(t+f\)
given the information available at time \(t\), can be expressed as:

\begin{equation}\label{eqn:lsr_prediction}
\begin{split}
& \E[_t]{y_{t+f}} =
    \E[_t]{c + \sum_{\tau = 1}^{F} \beta_{\tau} \tx_{t+f-\tau} + \epsilon_{t+f}} \\
    & = \E[_t]{c} + \sum_{\tau = 1}^{F} \E[_t]{\beta_{\tau} \tx_{t+f-\tau}} +
        \E[_t]{\epsilon_{t+f}} \\
    & = c + \sum_{\tau = 1}^{F} \beta_{\tau} \E[_t]{\tx_{t+f-\tau}}
\end{split}
\end{equation}

Sample estimates for \(c\) and \(\beta_{\tau}\) can be derived by
fitting an LSR model to the data available at time \(t\). However,
estimates for \(\tx_{t+f-\tau}\) at time \(t\) are only available for
\(\tau \geq f\). This leaves the task of estimating
\(\E[_t]{\tx_{t+f-\tau}}\) for \(\tau < f\), i.e., the future values of
the latent explanatory variable not yet known at the time of the
forecast.

The simplest approach would be to assume that \(\tx_t\) is i.i.d, so
that \(\E[_t]{\tx_i} = \E[_t]{\tx_j} = \E{\tx}\) for all \(i,j > t\).
In that case, \(\E{\tx}\) could be approximated by the sample mean of
\(\tx\). However, this approach risks discarding valuable information
about the future values of \(\tx\) which may be available from other
sources. A much more complex approach would be to try and predict the
future values of \(\tx\) using a separate lead-lag regression, but this
may add an unnecessary layer of complexity into the forecasting process.
This paper chooses the middle ground of assuming that \(\tx\) is not
i.i.d, but rather exhibits some first-order autocorrelation, namely:

\begin{equation}\label{eqn:lsr_shock_forecast}
\begin{split}
& \E{\tx_{t+f-\tau}} = \rho^{f-\tau} \, \tx_t +
                   \left(1 - \rho^{f-\tau} \right) \E{\tx} \\
& \text{for} \quad \tau < f
\end{split}
\end{equation}

where \(\rho\) is the first-order autocorrelation coefficient of \(\tx\)
and \(\E{\tx}\) is approximated by the sample mean of \(\tx\) at time
\(t\). This results in one additional coefficient having to be estimated
for each LSR model, increasing the total number of parameters from \(1 +
n + F\) to \(2 + n + F\) per forecast profile.

In addition to its relative simplicity, this choice of methodology has a
theoretical basis in past Phillips curve literature. For example,
inflation expectations are modelled as an AR(1) process in
\citep{Hommes-1998} and \citep{Hommes-2023}, while economic slack is
modelled as an AR(1) process in \citep{Woodford-2011} and, subsequently,
in \citep{Casarin-2025}.
\section{Empirical Results}
\label{empirical_results}
The main goal of this paper is to test the latent variable Phillips
curve (LVPC) hypothesis with reference to monetary policy applications.
To this end, the main empirical question addressed in this section is
whether a randomly selected Phillips curve factor specification would
generate more accurate medium-term inflation forecasts if implemented as
a traditional or a latent variable Phillips curve model. Performance
benchmarks are calculated using 3,968 combinatorially generated Phillips
curve factor specifications representing traditional, new Keynesian and
factor-augmented Phillips curves consistent with prior empirical PC
literature. Like-for-like model comparisons are proiritised over any
attempts to optimise for the accuracy of the individual Phillips curve
forecasts; in other words, no attempts are made to eliminate the
worst-performing models from the candidate pool, or to adjust the
individual factor specifications on a rolling basis based on in-sample
statistics such as parameter significance or goodness-of-fit.

As in the majority of prior empirical PC literature, the tests are
performed using historical evidence from the United States. Rolling OOS
forecasts are attempted on core Personal Consumption Expenditure (PCE)
inflation using quarterly data between Q1 1983 and Q1 2025. Observations
before Q1 1983 are omitted due to the structural break in the inflation
calculation methodology pointed out by, e.g., \citep{Stock-1999} and
\citep{Stock-2007}. The unit root hypothesis has not been rejected for
inflation in prior research, so the dependent variable used in all tests
is the change in sequential inflation, i.e., a twice-differenced natural
logarithm of the value of the core PCE price basket. The main results
are replicated for headline PCE inflation and discussed in Section
\ref{empirical_results_headline}.

The set of explanatory variables comprises eight alternative measures of
economic activity or slack (two labour market measures and six economic
activity measures selected from the set of predictors examined in
\citep{Stock-1999}), as well as four control variables: inflation
expectations (e.g., \citep{Albuquerque-2017}), the oil price (e.g.,
\citep{Salisu-2018,Chen-2014}), financial conditions (e.g.,
\citep{Conti-2021}), and the shadow federal funds rate defined in
\citep{Wu-Xia-2016}. For most of the explanatory variables, more than
one order of integration is considered, e.g., the output gap and the
change in the output gap are included as two standalone regression
candidates. 2,048 unique factor combinations are constructed by
combining each individual measure of economic activity with each
possible subset of control variables, including the empty set and the
full set. In an alternative specification, Atlanta Fed's sticky and
flexible CPI measures (\citep{Bryan-2010}) are used in lieu of inflation
expectations and oil prices, respectively, resulting in another 1,920
unique factor combinations. A full description of the input variables
and the underlying datasets can be found in tables
\ref{table:index_of_variables} and \ref{table:index_of_tickers}.

Forecasts are made between 1 and 8 quarters ahead. Mean Square
Prediction Error (MSPE) is used as the measure of forecast accuracy. A
minimum of 40 in-sample degrees of freedom (df)\footnote{The number of degrees of freedom is defined as the number of
available sample observations less the number of coefficients estimated
for a single in-sample model fit.} is set as a
prerequisite for producing an OOS forecast. This corresponds to a
minimum of 10 years of data on top of one data point lost for every
model coefficient. Exponentially decaying sample weights with a
half-life of 10 years are used to control for possible changes in the
shape of the Phillips curve over time. Point-in-time data is used for
all datasets for which version control is available. Due to the
differences in data release schedules for economic indicators with
point-in-time version control, data releases are checked twice per
quarter and a new forecast profile is generated if at least one new data
release is available. A forecast is only attempted if the latest
available data for all explanatory variables is at least as recent as
the latest data for the dependent.
\subsection{Traditional and LSR-based PC models}
\label{empirical_results_1}
The latent variable Phillips curve hypothesis states that the accuracy
of Phillips curve forecasts can be improved by treating the individual
explanatory variables in an empirical Phillips curve model as facets of
a single unobserved price pressure process. We can test this hypothesis
by comparing the accuracy of traditional ARX-based PC forecasts with the
accuracy of LSR-based (latent variable) PC forecasts using like-for-like
factor specifications. For each PC factor specification, forecasts are
made using five alternative methodologies: ARX(1), ARX(2), ARX(3),
ARX(4), and LSR. ARX forecasts are made using the multi-regression
approach specified in equation (\ref{eqn:lead_lag_multi_direct}), in order to
increase the number of degrees of freedom available at the start of the
forecast coverage and to avoid the need to infer future values for the
explanatory variables. For comparability with the ARX methodologies, LSR
models are specified with an autoregressive term.

The first Phillips curve forecasts are available as of 1999. The first
PC forecasts involving the oil price start in 2002. Some forecast
profiles are shorter because the corresponding models contain more
parameters and hence require more observations to satisfy the
requirement on the degrees of freedom. The distribution of model
coverage is summarised in Table \ref{table:model_coverage}.

\begin{figure*}[ht]
    \caption{MSPE rankings, like-for-like factor specifications: Core PCE}
    \includegraphics[width=\textwidth]{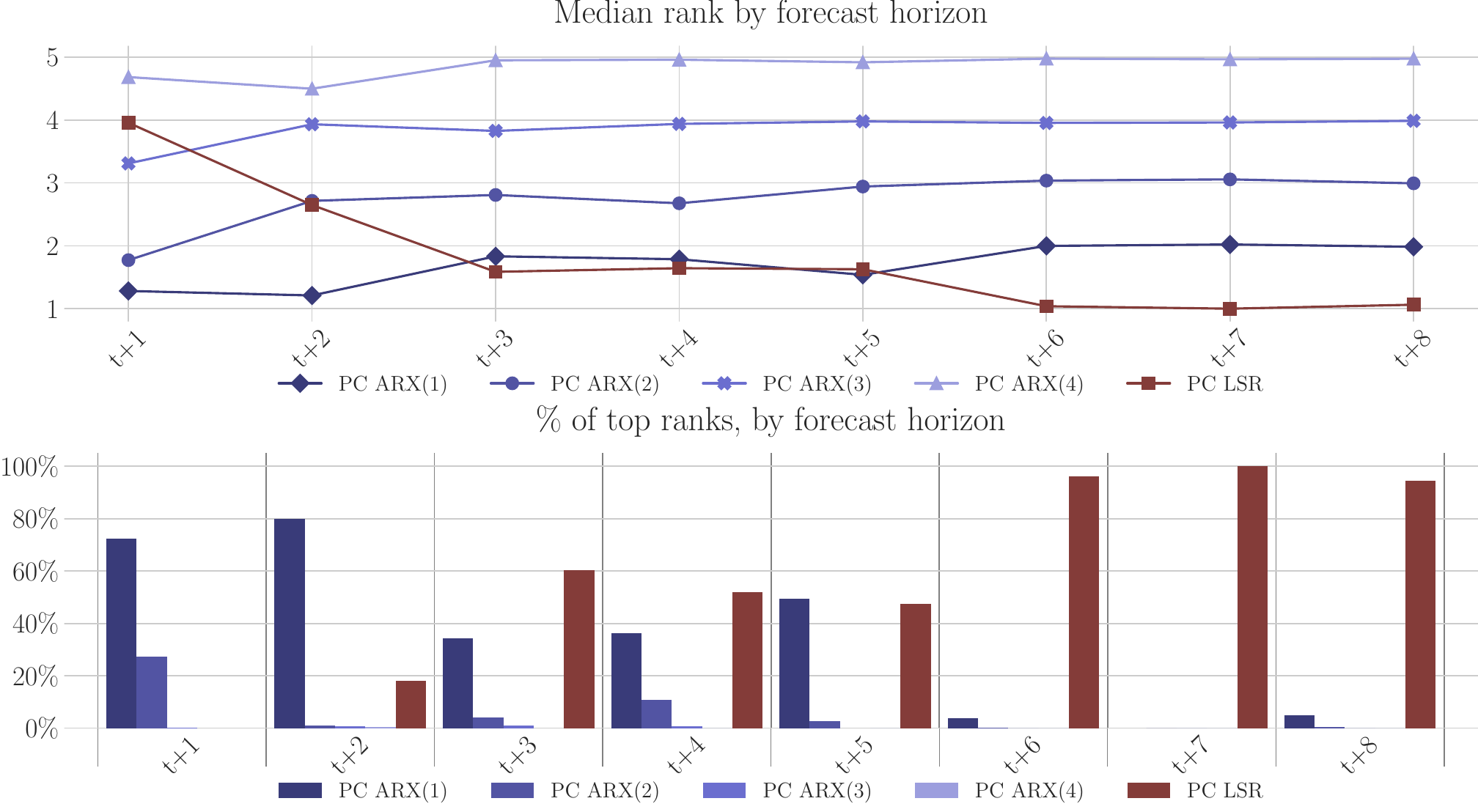}
    \label{fig-mspe_like_for_like_core}
\end{figure*}

For each factor combination, rolling eight-quarter forecast profiles are
generated for the full available history using each of the five
candidate methodologies. For each forecast horizon, the methodology that
produced the lowest MSPE is assigned a rank of 1 (highest accuracy), and
the methodology with the highest MSPE is assigned a rank of 5 (lowest
accuracy). This results in a total of \(3,968 \times 8 = 31,744\) ranks
from 1 to 5 for each candidate methodology. In Figure
\ref{fig-mspe_like_for_like_core}, the top chart shows the median rank
of each methodology at each forecast horizon, while the bottom chart
reports the percentage of factor specifications for which the given
methodology achieves a rank of 1.

For shorter-term inflation forecasts (1-2 quarters ahead), the
traditional ARX methodology seems to be the better choice. However, for
medium-term forecasts (6-8 quarters ahead) the LSR methodology is the
clear winner. For quarters that fall in-between, performance is roughly
similar with LSR models jumping to the top of the ranks in quarter 3 and
ARX models making a slight comeback in quarters 4 and 5. Within the ARX
family, the most parsimonious ARX model (order 1) predominantly
outperforms, suggesting that the higher-order ARX models overfit the
data. There is also some evidence of residual annual seasonality in core
PCE inflation data\footnote{Some evidence of seasonality in core PCE is reported in
\citep{Hornstein-2024}. The author concludes that the seasonal pattern
is much weaker after 2008.}, with ARX(1) and ARX(2) models gaining in
accuracy in quarters 4 and 5 and then again in quarter 8.

The overwhelming outperformance of LSR-based PC forecasts over the
medium term is promising. However, estimates of the aggregate price
pressure process calculated with 8 lags of data seem to produce
sub-optimal inflation forecasts over shorter horizons. This may be due
to seasonality and other statistical patterns in the inflation data
itself, or perhaps some degree of model misspecification in the current
set of PC factor combinations (this possibility is addressed in more
detail in Section \ref{model_stability}). A more worrisome possibility,
however, is that none of the PC models examined above contain any useful
information about the dynamics of inflation over the medium term, and
LSR-based PC models only outperform by being marginally less overfitted
and hence generating predictions that are closer to a naïve model.

\begin{figure*}[ht]
    \caption{MSPE rankings incl. univariate models and EWMA: Core PCE}
    \includegraphics[width=\textwidth]{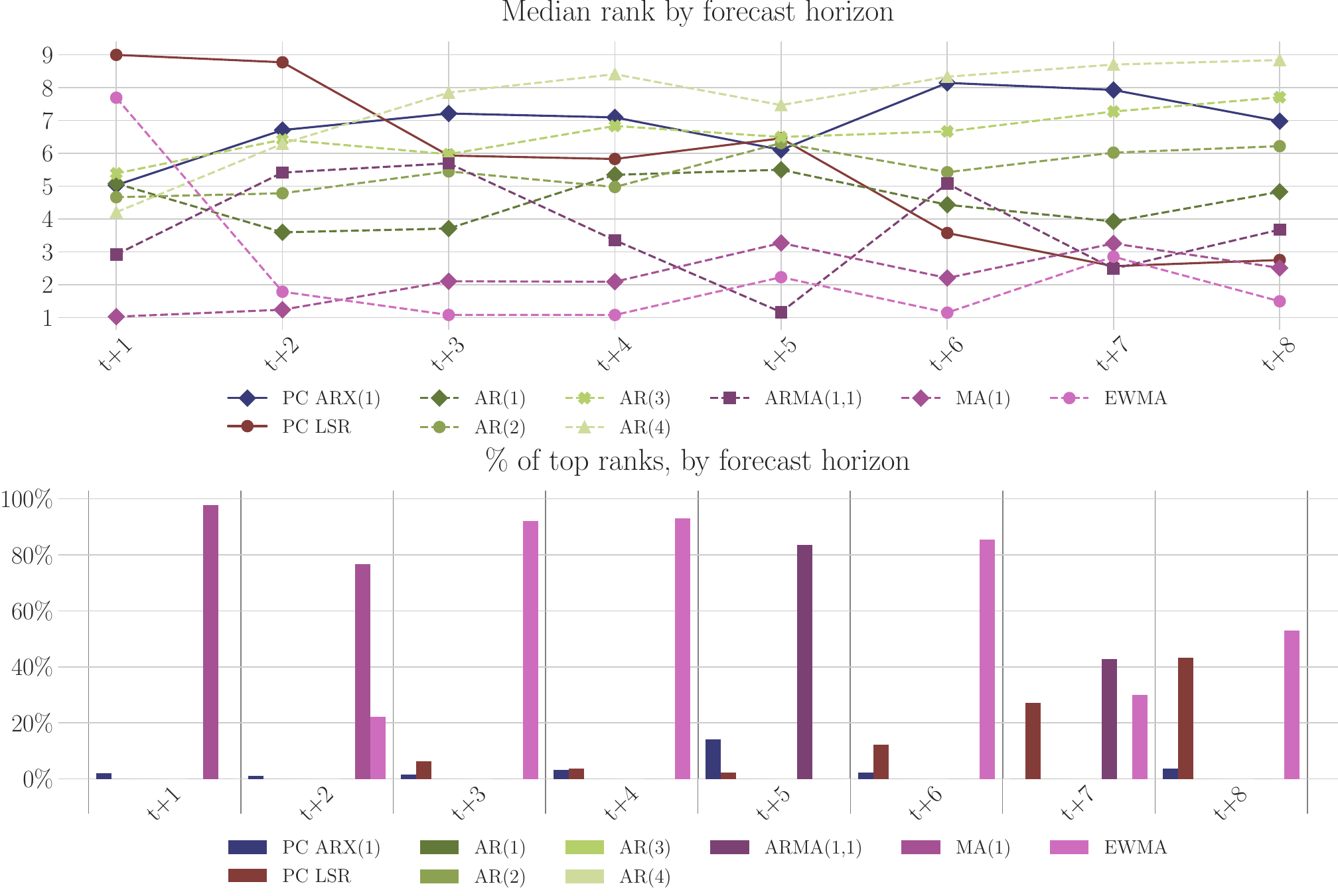}
    \label{fig-mspe_ranks_all_core}
\end{figure*}

In order to address this possibility, we can include a set of univariate
benchmarks in the model comparison. Figure \ref{fig-mspe_ranks_all_core}
expands the model rankings to include seven univariate inflation models
from prior empirical PC literature: a rolling sample mean model,
autoregressive models of orders AR(1) to AR(4), a moving average model
of order MA(1), and an autoregressive moving average model of order
ARMA(1,1). Note that the rolling sample mean model is technically an
exponentially weighted moving average (EWMA) model due to exponentially
decaying sample weights. In each case, univariate forecasts are
truncated to have the same historical coverage as the respective PC
model. Higher-order ARX models are excluded as they underperform the
ARX(1) model for most specifications. As before, a rank of 1 (9) is
assigned to the methodology which produces the lowest (highest) MSPE for
a given PC factor combination at a given forecast horizon.

The best methodology for one-quarter-ahead inflation forecasts is now a
simple MA(1) model, confirming the conclusion from \citep{Stock-2007}
that core PCE inflation is IMA(1,1). The ARMA(1,1) and ARX(1) models
outperform at the 5-quarter mark, once again pointing to some degree of
annual seasonality in the inflation data.

The Phillips curve models as a group do not fare particularly well
against the MA(1) and EWMA models\footnote{Beyond the first quarter, MA(1) forecasts are very similar to
EWMA forecasts by design} -- a finding consistent with
prior literature; however, a meaningful and increasing share of
LSR-based PC models outperform all univariate benchmarks as the horizon
extends beyond 5 quarters: 12\% in quarter 6, 27\% in quarter 7, and 45\%
in quarter 8. This suggests that a latent variable Phillips curve model
can provide useful information about the medium-term dynamics of core
PCE inflation provided that the choice of factor combination is not
entirely random. The same does not hold for traditional ARX(1)-based PC
models whose chance of outperforming a univariate benchmark over six to
eight quarters stands at a negligible 2\%, 0\% and 4\%, respectively.
\subsection{PC Models with MA(1) Residuals}
\label{empirical_results_2}
Results from the previous section confirm the finding from
\citep{Campbell-2007} that changes in sequential core PCE inflation
follow an MA(1) process. A natural question is whether incorporating
MA(1) residuals into empirical PC models would affect the conclusions
from the previous section. To address this question, a new set of OOS
forecasts is made using the following PC methodologies:

\begin{enumerate}
\item X(1), LSR-AR(0): Traditional and LSR-based PC models which exclude
the autoregressive lag of inflation and only incorporate the
explanatory variables.
\item ARX(1), LSR: Traditional and LSR-based PC models which incorporate
one autoregressive lag of inflation -- same as in the the previous
section.
\item MAX(1), LSR-MA(1): Models with an MA(1) innovation process for the
residuals but no autoregressive lags of inflation.
\item ARMAX(1,1), LSR-ARMA(1,1): Models with one autoregressive lag of
inflation and an MA(1) innovation process for the residuals.
\end{enumerate}

For models with an MA(1) component, initial parameters are calculated
using the Hannan-Rissanen method (\citep{hannan-rissanen-1982}).
Gaussian maximum likelihood optimisation is then performed using the
Nelder-Mead algorithm (\citep{nelder-mead-1965}) with a hard limit of
10,000 iterations to ensure reasonable computation times. The LSR
methodology is adjusted for MA(1) residuals by modifying equation
(\ref{eqn:lsr}) in the following way:

\begin{equation}\label{eqn:lsr_ma}
y_t = c + \sum_{\tau = 1}^{F} \beta_{\tau} \tx_{t-\tau} + \epsilon_t + \theta \epsilon_{t-1}
\end{equation}

\begin{figure*}[ht]
    \caption{MSPE rankings after MA(1) adjustment of PC models: Core PCE (1)}
    \includegraphics[width=\textwidth]{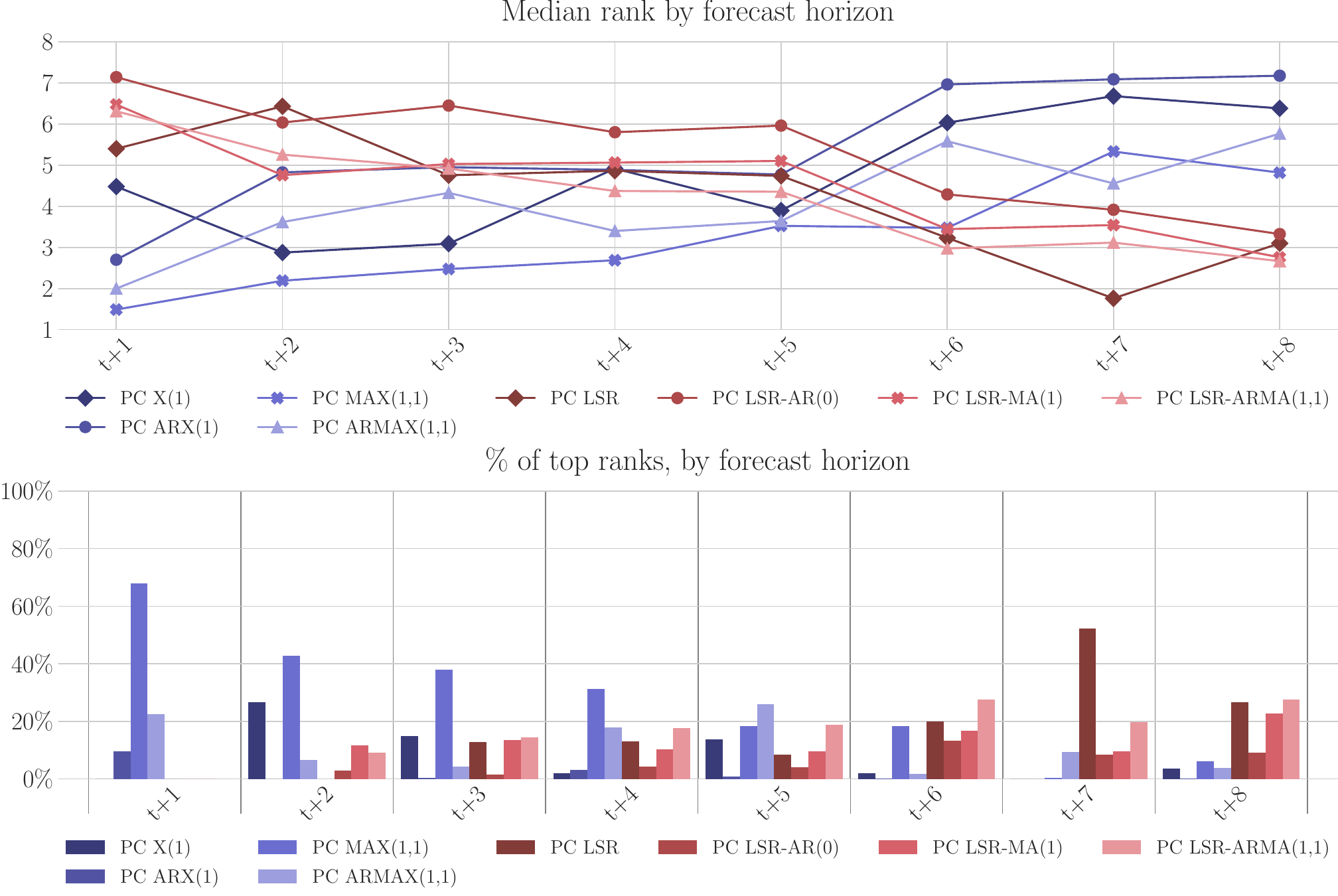}
    \label{fig-mspe_like_for_like_core_ma}
\end{figure*}

Figure \ref{fig-mspe_like_for_like_core_ma} summarises the relative
performance of the new set of methodologies using the same approach as
in Figure \ref{fig-mspe_like_for_like_core}. The results confirm that
LSR-based PC forecasts are more accurate than the traditional PC
forecasts over the medium term (i.e., around 2 years ahead).
Furthermore, methodologies that include an MA(1) adjustment for the
residuals generally outperform those without the adjustment. The
improvement is consistent across all forecast horizons for traditional
PC models. MAX(1,1)-based models outperform up to 4 quarters ahead, and
only slightly underperform the ARMAX(1,1)-based models in quarter 5 --
once again likely due to an annual seasonality pattern in inflation. The
improvement from including MA(1) residuals in LSR models is somewhat
less pronounced: the simple LSR methodology still outperforms in quarter
7 and performs at par with the LSR-ARMA(1,1) model in quarters 3 and 8.
This may, however, be due to technical reasons, as some of the
MA(1)-augmented LSR models struggle to converge under the hard limit of
10,000 iterations used in this study.

\begin{figure*}[ht]
    \caption{MSPE rankings after MA(1) adjustment of PC models: Core PCE (2)}
    \includegraphics[width=\textwidth]{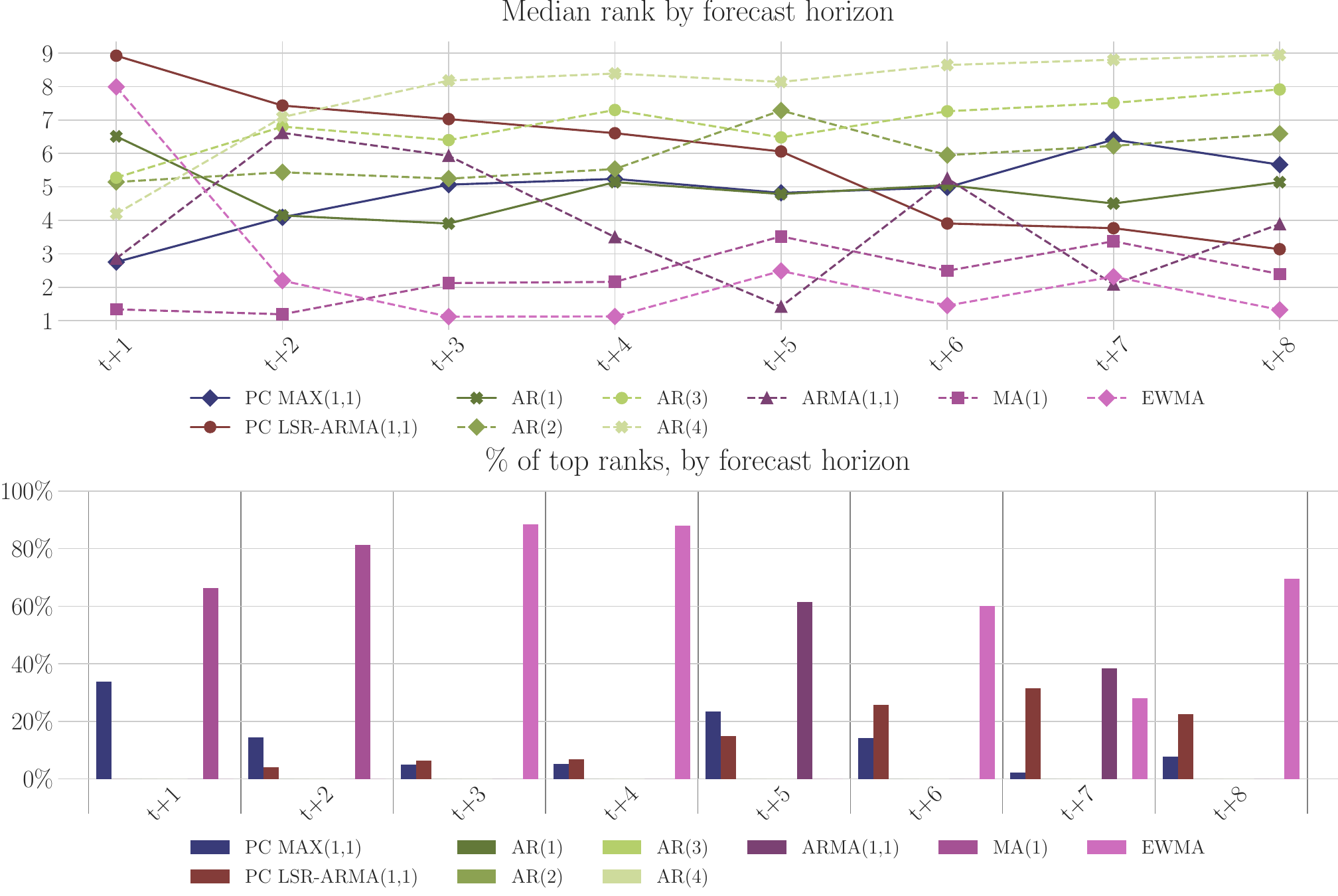}
    \label{fig-mspe_ranks_all_core_ma}
\end{figure*}

Forecasts from the MAX(1,1) and LSR-ARMA(1,1) PC models are compared
against the univariate benchmarks in Figure
\ref{fig-mspe_ranks_all_core_ma}. 34\% of MAX(1,1) PC models now
outperform all univariate benchmarks in quarter 1, and 25\% outperform in
quarter 5. As before, the share of LSR-ARMA(1,1) PC models with
best-in-class forecasts increases steadily with the time horizon, from
4\% in quarter 2 to 32\% in quarter 7, only dropping slightly to 24\% for
quarter 8. This suggests, once again, that a latent variable Phillips
curve model can offer useful information about the dynamics of inflation
over the medium term if its specification is not selected at random.
\subsection{Statistical Significance and Model Stability}
\label{model_stability}
The tests performed in this paper so far have been primarily concerned
with the \emph{likelihood} of a randomly chosen PC model outperforming a
univariate benchmark. However, an equally important question is the
\emph{magnitude} of the improvement.

\begin{figure*}[ht]
\caption{Statistical significance of model predictions after MA(1) adjustment: Core PCE}
\includegraphics[width=\textwidth]{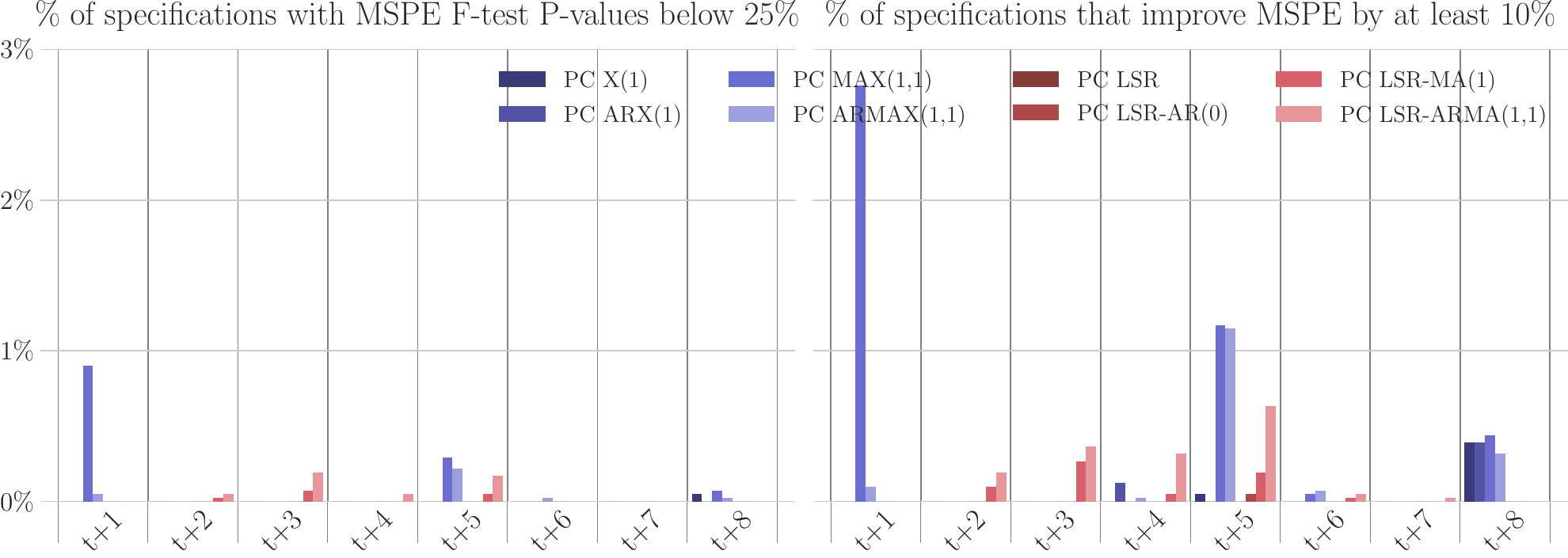}
\label{fig-f_tests_core_ma}
\end{figure*}

In Figure \ref{fig-f_tests_core_ma}, the chart on the left shows the
percentage of PC specifications whose MSPE improvement over the MA(1)
model is significant at the 25\% level\footnote{None of the models examined in this study live up to the standard
5\% significance threshold.} using a standard one-sided
F-test. The chart on the right reports the percentage of specifications
whose MSPE is at least 10\% lower than that of the MA(1) model. As
before, the results are broken down by methodology and by forecast
horizon. Note that the vertical axis is scaled to a maximum of 3\%,
corresponding to approximately 120 out of the 3,968 combinatorial factor
specifications in the study.

Although a non-negligible share of PC models outperform the MA(1)
benchmark by a small margin, a negligible percentage of these models
offer a \emph{statistically significant} improvement, even at the 25\%
significance level, and a very small percentage improve on the MA(1)
MSPE by more than 10\%. These results are loosely consistent with prior
research, and they are not entirely unexpected given this paper's design
choice to prioritise like-for-like methodology comparisons over
case-by-case forecasting accuracy. Still, some discussion of the
underlying reasons is in order, and one possible explanation lies in the
area of model stability.

Firstly, it is well documented in prior academic literature that
inflation is prone to bouts of volatility in the aftermath of shock
events (e.g., see \citep{Lansing-2009,Eisenstat-2016}). The presence of
these episodes in the data sample may disproportionately affect both
model calibration and model selection when standard variance-based
approaches are used to measure goodness of fit. For a simple visual
confirmation, we can examine the MA(1)-adjusted PC models that achieve
the best ex-post MSPE rankings in the current study. Table
\ref{table:best_factor_groups_core_ma} reports the best MSPE improvement by
forecast horizon, while Figure \ref{fig-best_models_by_horizon_core_ma} plots
the corresponding models' actual real-time predictions of inflation.
Based on a simple visual inspection, all of the best-performing models
tend to do reasonably well during the post-COVID inflation surge of
2021-2023; however, the performance of these models is far less uniform
elsewhere. In other words, the standard model calibration practice of
using long-term inflation data at face value may inadvertently favour
the models which perform well during crisis periods, even if their
performance is not consistently strong in normal times\footnote{The distinction between PC models optimised for ``normal'' and
``crisis'' regimes is an important one in light of the evidence from
prior literature. For example, \cite{Ball-1988} derives a new Keynesian
Phillips curve model with price rigidities, concluding that high
inflation and inflation volatility regimes lead to changes in the
producers' price setting behaviour. Subsequently, \cite{Ball-2011} and
\cite{Mutascu-2019}, among others, provide empirical evidence that the
Phillips curve changes shape during crises and downturns.}.

The second important factor in model stability is the effect of variable
selection and the risk of misspecification. We can gauge this risk by
examining the effects of including or omitting individual variables from
the pool of candidate regressors, taking the set of 3,968 LSR-ARMA(1,1)
LVPC models as an example\footnote{The results for MAX(1,1) models are largely similar and omitted
here for brevity.}. First, a global median MSPE
(\(MSPE_{all}\)) is calculated across all factor specifications for each
forecast horizon. Then, for each explanatory variable in the candidate
pool, the median MSPE is calculated for all factor specifications in
which the variable is present (\(MSPE_{var}\)). The MSPE effect of
including a variable is then calculated as:

\begin{equation*}
\frac{MSPE_{var} - MSPE_{all}}{MSPE_{bm}}
\end{equation*}

where \(MSPE_{bm}\) is the MSPE of the MA(1) model.

The results are reported in Figure \ref{fig-mspe_predictors_core_ma}. Bars
below the zero line indicate an improvement in forecast accuracy (the
median model which contains the variable in question has a lower MSPE
than the global median model for the given forecast horizon). Bars above
the zero line indicate a deterioration.

No single explanatory variable improves the median forecasting accuracy
uniformly across horizons. The change in capacity utilisation looks like
a reasonably strong predictor of inflation, except for the horizons
corresponding to annual inflation seasonality patterns (quarters 4, 5
and 8). This result is consistent with the findings from
\citep{Stock-1999}, which identifies capacity utilisation and
manufacturing and trade sales as two of the best-performing variables in
empirical PC models. Otherwise, the strongest medium-term predictors of
inflation (6 to 8 quarters ahead) seem to be the unemployment gap and
the (nominal and real) output gap, while the strongest shorter-term
models tend to include sticky and/or flexible price CPI inflation and
nominal or real GDP growth. In fact, models with real GDP growth and
sticky price CPI inflation tend to have a higher median forecast
accuracy across all forecast horizons, but improvement over the medium
term is marginal at best.

The implications of the results in Figure \ref{fig-mspe_predictors_core_ma}
are different from the perspective of the traditional Phillips curve and
the latent variable Phillips curve. In traditional Phillips curve
models, variables like unemployment or the output gap are assumed to
influence future inflation directly. From this viewpoint, the difference
in lag profiles may imply that inflation is influenced by different
macroeconomic forces over different time horizons. In contrast, latent
variable Phillips curve models view inflation as a staggered reflection
of a singular unobserved variable. This, in turn, implies that there
exists exactly one ``correct'' lag profile of transmission, and any
deviation from this lag profile in the observed data can be viewed as
measurement error\footnote{Recall that in LSR models, all explanatory variables affect the
dependent with a shared lag profile.}. In other words, heterogenous lag profiles imply
that the set of regression candidates considered in this study is a
sub-optimal one for a latent variable Phillips curve model.
\subsection{Headline PCE Inflation}
\label{empirical_results_headline}
Figures
\ref{fig-mspe_like_for_like_core}-\ref{fig-mspe_ranks_all_core_ma} are
replicated for headline PCE inflation in Figures
\ref{fig-mspe_like_for_like_headline}-\ref{fig-mspe_ranks_all_headline_ma}
in the Appendix. The core PCE basket excludes the most volatile
components of inflation, namely, food and energy prices, which are
included in the headline PCE measure. This makes headline PCE inflation
more volatile and more susceptible to short-term economic fluctuations.
In theory, this should imply that headline PCE is more predictable in
the short run but more difficult to predict over longer periods.

That being said, the results for headline PCE inflation are generally
similar to those for core PCE. Once again, LSR-based PC models
outperform in medium-term forecasts. Like with core, some evidence of
annual seasonality exists in the headline inflation measure, with the
ARMA(1,1) model dominating in quarters 1 and 5 and the AR(1) model
beating 60\% of the PC models in quarter 8.

PC models without the MA(1) adjustment generally fail to outperform any
univariate benchmarks. However, MAX(1,1)-based PC models stand a
comparable chance of outperforming a univariate benchmark for headline
PCE inflation as for core, and LSR-ARMA(1,1) models once again steadily
gain in forecasting accuracy over longer horizons with an 18\% and 21\%
chance of outperforming all univariate benchmarks in quarters 7 and 8,
respectively. Unlike with core PCE, almost none of the PC models produce
a statistically or economically significant improvement in MSPE for
headline PCE inflation beyond one period ahead, likely owing to the more
volatile and more myopic nature of the headline PCE measure.
\section{Concluding Remarks}
\label{concluding_remarks}
The main goal of this paper was to examine the merits of the latent
variable Phillips curve (LVPC) hypothesis based on the performance of
LVPC models in medium-term inflation forecasting. Strong evidence was
found that LVPC models implemented with the Latent Shock Regression
(LSR) methodology outperform traditional ARX-based PC models when
forecasting US core PCE inflation six to eight quarters ahead. This
outperformance is not fully explained by the lower propensity of LSR
models to overfit the data: In a sample of 3,968 Phillips curve factor
specifications, a randomly selected factor combination implemented as a
latent variable PC model stands a 45\% chance of beating the best
univariate benchmark eight quarters ahead, compared to a 4\% chance for a
traditional ARX-based PC model.

As previously mentioned, this paper makes no explicit attempt to
identify the ``best'' empirical PC model of inflation. The pool of PC
factor combinations is generated at random with no prejudice about the
forecasting power of the individual predictors. No further attempts are
made to optimise the performance of the individual PC models through
real-time feature selection or other techniques. As an expected side
effect, the performance gains achieved by the PC models examined in this
study are statistically are economically small compared to the standard
univariate benchmarks. This leaves much scope for further performance
improvement in future empirical work, and two sets of relevant
observations can be made to this end.

The first set of observations is in regards to the statistical
properties of inflation itself. As a starting point, controlling for an
IMA(1,1) process in the inflation data, as proposed by
\citep{Stock-2007}, makes PC forecasts more competitive against the
standard univariate benchmarks. This result is corroborated by the
out-of-sample performance of MA(1)-augmented PC models reported in
Section \ref{empirical_results_2}. This paper's empirical results also point
to annual seasonality patterns in core and, to some extent, in headline
PCE inflation. Although the assumption of seasonality in the inflation
data has been empirically challenged for the more recent releases (e.g.,
see \citep{Hornstein-2024}), controlling for seasonal patterns may still
be worthwhile for PC models which are calibrated using a long data
history. Finally, persistent spikes in inflation volatility after shock
events may introduce a bias into the model optimisation process,
favouring accuracy during crisis periods over accuracy in normal times.
This may be problematic because the Phillips curve has been shown to
change shape during crisis regimes (e.g., \citep{Mutascu-2019,Ball-2011,Ball-1988}).

The second set of observations concerns the choice of regression
candidates for empirical Phillips curve models. This paper's findings
suggest that treating the traditional Phillips curve factors as proxies
of a singular unobserved price pressure process improves the accuracy of
inflation forecasts over the medium term. However, as discussed in
Section \ref{model_stability}, this does not imply that traditional Phillips
curve factors are the best proxies for this unobserved price pressure
process to begin with. For example, if price pressure is related to
capacity utilisation, then its dynamics may be captured by fluctuations
in the equity market before the official capacity utilisation data are
released by the Federal Reserve System. Likewise, survey-based measures
of business activity may contain timelier information about the real
economy as compared to the more backward-looking, revision-prone
quarterly aggregates such as unemployment and GDP.

Bringing all these observations together, the following recommendations
can be made for improving the competitiveness of empirical PC models in
real-life inflation forecasting settings. First, competitive PC models
of inflation need to account for a moving average process in the error
term. Second, PC models calibrated with a long data history may need to
control for residual annual seasonality. Third, medium-term PC forecasts
should be implemented using the LSR methodology and the search for
candidate regressors expanded to include more homogenous and
forward-looking explanatory variables such as business activity surveys
and financial market aggregates. Fourth, PC models designed for
modelling inflation in normal times (as opposed to crisis times) may
need to exclude periods of high inflation volatility such as the
post-COVID period of 2021-2023, or at least incorporate some form of
heteroskedasticity adjustment into the regression design. These
methodological improvements should help to further advance the Phillips
curve as a practical tool for real-life inflation forecasting in future
empirical work.

\appendix
\section{Derivation of the LSR Fixed Point Using the Method of Least Squares}
\label{lsr_derivation}
Assume WLOG that equation (\ref{eqn:lsr_vec}) is being estimated using a data
sample with \(s\) observations. The least squares objective function in
sample form is then given by:

\begin{equation}\label{eqn:lsr_optim_problem} 
\underset{c,\B,\o}{\min} \, \| \y - \1_s c - \P \bo \|^2_2
\end{equation}

where \(\y\) is an \(s \times 1\) sample observation vector for \(y\),
\(\P\) is an \(s \times nF\) sample observation matrix for \(P\), and
\(\1_s\) is an \(s \times 1\) vector of ones. The solution for the
intercept term \(c\) is both well documented and derived in
\citep{Bargman-clarx-2025-preprint}, and is given by:

\begin{equation}\label{eqn:lsr_solution_c}
c = \overline{\y} - \overline{\P} \bo
\end{equation}

where \(\overline{\y}\) is the sample mean of \(\y\) and
\(\overline{\P}\) is the column-wise sample mean of \(\P\). Plugging the
solution for \(c\) back into (\ref{eqn:lsr_optim_problem}) and expanding, we
obtain:

\begin{equation}\label{eqn:lsr_optim_short} 
\underset{c,\B,\o}{\min} \, \Sm_{y} - 2 \,  \bo' \Sm_{Py} + \bo' \Sm_{P} \bo
\end{equation}

where \(\Sm_{y}\) is the sample variance of \(y\), \(\Sm_{P}\) is the
sample covariance matrix of \(P\), and \(\Sm_{Py}\) is the sample
covariance from \(P\) to \(y\).

The objective function given by (\ref{eqn:lsr_optim_short}) is a convex
optimisation problem which can be solved by setting its partial
derivatives to zero. We also note that by the properties of the
Kronecker product, the term \(\bo\) can be factorised as follows:

\begin{equation*} 
\bo = \bio = \oib
\end{equation*}

Substituting \(\bo\) for \(\bio\) and taking a partial derivative of
(\ref{eqn:lsr_optim_short}) with respect to \(\o\) yields:

\begin{align*}
& \frac{\partial}{\partial \, \o } \, \left[ \Sm_{y} - 2 \,  \bo' \Sm_{Py}
               + \bo' \Sm_{P} \bo \right] = \\[10pt]
= \, & \frac{\partial}{\partial \, \o } \, \left[ \o' \bi' \Sm_{P} \bio
               - 2 \, \bio' \Sm_{Py} \right] = \\[10pt]
= \, & 2 \, \bi' \Sm_{P} \bio - 2  \,  \bi' \Sm_{Py}
\end{align*}

Setting to zero and solving for \(\o\), we obtain:

\begin{equation*}
\o = \left[ \bi' \Sm_{P} \bi \right]^{-1} \bi' \Sm_{Py}
\end{equation*}

By symmetry, the solution for \(\B\) is is given by:

\begin{equation*}
\B = \left[ \oi' \Sm_{P} \oi \right]^{-1} \oi' \Sm_{Py}
\end{equation*}

\qed

\pagebreak
\section{Charts and tables}
\label{charts_and_tables}
\vspace{10pt}

\begin{table*}[ht]
\centering
\caption{Data series used in empirical Phillips curve modelling}
\label{table:index_of_tickers}
\begin{adjustbox}{width=\textwidth, nofloat=table}
\begin{threeparttable}
  \begin{tabular}{lllll}
  \toprule
  Data series & Source & History start & Frequency & Version control \\
  \midrule
  Average Hourly Earnings & Bureau of Labor Statistics & 1964-01 & Monthly & as of 1999-08-06 \\
  Brent crude oil price & U.S. Energy Information Administration & 1987-05-20 & Daily & None / Disabled \\
  Capacity utilisation & Federal Reserve System & 1967-01 & Monthly & as of 1996-11-15 \\
  Chicago Fed ANFCI & Federal Reserve Bank of Chicago & 1971-01-04/1971-01-10 & Weekly & None / Disabled \\
  Cleveland Fed infl. exp. (2y) & Federal Reserve Bank of Cleveland & 1982-01 & Monthly & as of 2021-10-13 \\
  Core PCE index & Bureau of Economic Analysis & 1959-01 & Monthly & as of 2000-08-01 \\
  Federal funds effective rate & Federal Reserve System & 1954-07 & Monthly & None / Disabled \\
  Flexible price CPI & Atlanta Fed & 1967-03 & Monthly & as of 2014-03-07 \\
  Headline PCE index & Bureau of Economic Analysis & 1959-01 & Monthly & as of 2000-08-01 \\
  Industrial production & Federal Reserve System & 1919-01 & Monthly & None / Disabled \\
  Natural rate of unemployment (NAIRU) & Congressional Budget Offie & 1949Q1 & Quarterly & as of 2011-02-02 \\
  Nominal GDP & Bureau of Economic Analysis & 1946Q1 & Quarterly & as of 1991-12-04 \\
  Nominal potential GDP & Congressional Budget Office & 1949Q1 & Quarterly & as of 2011-02-02 \\
  Real GDP & Bureau of Economic Analysis & 1947Q1 & Quarterly & as of 1991-12-04 \\
  Real potential GDP & Congressional Budget Office & 1949Q1 & Quarterly & as of 1991-01-30 \\
  Sticky price CPI & Atlanta Fed & 1967-01 & Monthly & as of 2014-03-07 \\
  Unemployment rate & Bureau of Labor Statistics & 1948-01 & Monthly & as of 1960-03-15 \\
  Wu-Xia shadow federal funds rate & Wu and Xia (2016) & 1990-01 & Monthly & None / Disabled \\
  \bottomrule
  \end{tabular}
  
\begin{tablenotes}[flushleft]
 \item Note: All data series except the Wu-Xia shadow federal funds
 rate (\citep{Wu-Xia-2016}) are available from the St. Louis
 Federal Reserve economic database (FRED/ALFRED). The Wu-Xia shadow
 federal funds rate was taken directly from the website of the
 Federal Reserve Bank of Atlanta.
\end{tablenotes}
\end{threeparttable}
\end{adjustbox}
\end{table*}


\begin{table*}[ht]
\centering
\caption{Variables used in empirical Phillips curve modelling}
\label{table:index_of_variables}
\begin{adjustbox}{width=\textwidth, nofloat=table}
\begin{threeparttable}
    \begin{tabular}{lllll}
    \toprule
    Variable & Data series\tnote{1} & Calculation steps & Differencing rules\tnote{2} & Used as \\
    \midrule
    Brent crude price \%-chg & Brent crude oil price & natural log, first difference & lvl, (d) & control \\
    Capacity utilisation & Capacity utilisation &  & lvl, (d) & factor \\
    Chicago Fed ANFCI & Chicago Fed ANFCI &  & (d), (dd) & control \\
    Cleveland Fed infl. exp. (2y) & Cleveland Fed infl. exp. (2y) &  & lvl, (d) & control \\
    Core PCE sequential inflation & Core PCE index & natural log, first difference & (d) & dependent \\
    Federal funds rate & Wu-Xia shadow federal funds rate, Federal funds effective rate & take second if first N/A & lvl, (d) & control \\
    Flexible price CPI & Flexible price CPI &  & lvl, (d) & control \\
    Headline PCE sequential inflation & Headline PCE index & natural log, first difference & (d) & dependent \\
    Ind. prod. sequential growth & Industrial production & natural log, first difference & lvl, (d) & factor \\
    Nominal GDP sequential growth & Nominal GDP & natural log, first difference & lvl, (d) & factor \\
    Output gap (nominal) & Nominal GDP, Nominal potential GDP & natural log, subtract second from first & lvl, (d) & factor \\
    Output gap (real) & Real GDP, Real potential GDP & natural log, subtract second from first & lvl, (d) & factor \\
    Real GDP sequential growth & Real GDP & natural log, first difference & lvl, (d) & factor \\
    Sticky price CPI & Sticky price CPI &  & lvl, (d) & control \\
    Unemployment gap (U3) & Unemployment rate, Natural rate of unemployment (NAIRU) & subtract second from first & lvl, (d) & factor \\
    Wage growth (AHE) & Average Hourly Earnings & natural log, first difference & lvl, (d) & factor \\
    \bottomrule
    \end{tabular}
    
\begin{tablenotes}[flushleft]
\item[1] All underlying data series are listed in table
\ref{table:index_of_tickers}. If the calculation of a variable involves
more than one series, the series names are separated by a comma.
\item[2] A list of differencing operations applied to each variable.
  `lvl' implies that the variable is included in the regression pool
  without a further differencing operation; `(d)' means
  once-differenced, `(dd)' twice-differenced, etc. All listed
  differencing levels for each variable are included as separate
  candidates in the regression pool. The differencing level is used
  as a suffix in the variable name in the final output (`lvl' is
  omitted).
\end{tablenotes}
\end{threeparttable}
\end{adjustbox}
\end{table*}

\begin{table*}[ht]
\footnotesize
\caption{Summary of PC model forecast coverage by year}
\label{table:model_coverage}
\begin{adjustbox}{width=\textwidth, nofloat=table}
\centering
\begin{threeparttable}
  \begin{tabular}{p{0.1\linewidth} p{0.8\linewidth} c}
  \toprule
  Year & New variables included in factor specifications & New models \\
  \midrule
  1999 & Chicago Fed ANFCI, Nominal GDP sequential growth, Unemployment gap (U3), Ind. prod. sequential growth, Federal funds rate, Wage growth (AHE), Flexible price CPI, Capacity utilisation, Output gap (nominal), Sticky price CPI, Real GDP sequential growth, Cleveland Fed infl. exp. (2y), Output gap (real) & 16 \\
  2000 &  & 96 \\
  2001 &  & 304 \\
  2002 & Brent crude price \%-chg & 544 \\
  2003 &  & 560 \\
  2004 &  & 336 \\
  2005 &  & 144 \\
  2006 &  & 192 \\
  2007 &  & 400 \\
  2008 &  & 480 \\
  2009 &  & 320 \\
  2010 &  & 112 \\
  2011 &  & 16 \\
  \bottomrule
  \end{tabular}
  
\begin{tablenotes}[flushleft]
\item Notes: 1. Years with no change in coverage are excluded. A
variable is shown in the table when it has accumulated enough data
history to be included in the first forecasting model, rather than
when its own data history starts. 2. Due to differences in factor
specifications and/or lag orders, forecast history for some models
starts at a later date than others to fulfill the the
degrees-of-freedom threshold. 3. The Employment Cost Index (ECI) was
included in the original pool of explanatory variables but was later
removed from the study. The ECI methodology was revised in 2006,
causing the point-in-time versions of the ECI series after 2006 to
have a shorter data history. As a result, some of the forecasting
models whose factor specifications include ECI have a temporary gap in
coverage which coincides with the 2008 crisis, making them less
comparable with other forecast specifications on OOS forecasting
power.
\end{tablenotes}
\end{threeparttable}
\end{adjustbox}
\end{table*}

\renewcommand{\arraystretch}{1.25}
\begin{table*}[ht]
  \fontsize{8}{10}
  \selectfont
  \centering
  \caption{Best-performing PC model by forecast horizon: Core PCE}
  \begin{adjustbox}{width=\textwidth, nofloat=table}
  \begin{threeparttable}
  \begin{tabular}{c c p{0.7\linewidth} c}
  \toprule
  Horizon & Methodology & Factor specification & Gain\tnote{1} \\
  \midrule
  t+1 & MAX(1,1) & Output gap (real) (d), Cleveland Fed infl. exp. (2y) (d), Federal funds rate (d), Brent crude price \%-chg (d) & 17.1\% \\
  t+2 & LSR-ARMA(1,1) & Real GDP sequential growth, Sticky price CPI, Sticky price CPI (d), Flexible price CPI, Federal funds rate (d), Chicago Fed ANFCI (dd) & 12.7\% \\
  t+3 & LSR-ARMA(1,1) & Ind. prod. sequential growth (d), Federal funds rate (d), Brent crude price \%-chg, Chicago Fed ANFCI (d), Chicago Fed ANFCI (dd) & 20.5\% \\
  t+4 & LSR-ARMA(1,1) & Ind. prod. sequential growth, Sticky price CPI (d), Flexible price CPI, Federal funds rate (d), Chicago Fed ANFCI (dd) & 15.9\% \\
  t+5 & ARMAX(1,1) & Output gap (real), Cleveland Fed infl. exp. (2y), Cleveland Fed infl. exp. (2y) (d), Federal funds rate (d), Brent crude price \%-chg, Brent crude price \%-chg (d), Chicago Fed ANFCI (d) & 20.0\% \\
  t+6 & ARMAX(1,1) & Output gap (real), Cleveland Fed infl. exp. (2y), Cleveland Fed infl. exp. (2y) (d), Brent crude price \%-chg, Brent crude price \%-chg (d) & 17.6\% \\
  t+7 & LSR-ARMA(1,1) & Output gap (real), Cleveland Fed infl. exp. (2y), Cleveland Fed infl. exp. (2y) (d), Federal funds rate (d), Brent crude price \%-chg, Chicago Fed ANFCI (dd) & 10.1\% \\
  t+8 & MAX(1,1) & Wage growth (AHE) (d), Cleveland Fed infl. exp. (2y), Cleveland Fed infl. exp. (2y) (d), Federal funds rate (d), Brent crude price \%-chg, Brent crude price \%-chg (d), Chicago Fed ANFCI (dd) & 15.3\% \\
  \bottomrule
  \end{tabular}
  
  \begin{tablenotes}[flushleft]
  \footnotesize
  \item[1] Improvement in MSPE compared to an MA(1) model.
  \end{tablenotes}
  \end{threeparttable}
  \end{adjustbox}
  \label{table:best_factor_groups_core_ma}
\end{table*}

\FloatBarrier

\begin{figure*}[ht]
    \caption{Best-performing MA(1)-augmented PC model by forecast horizon: Core PCE}
    \includegraphics[width=\textwidth]{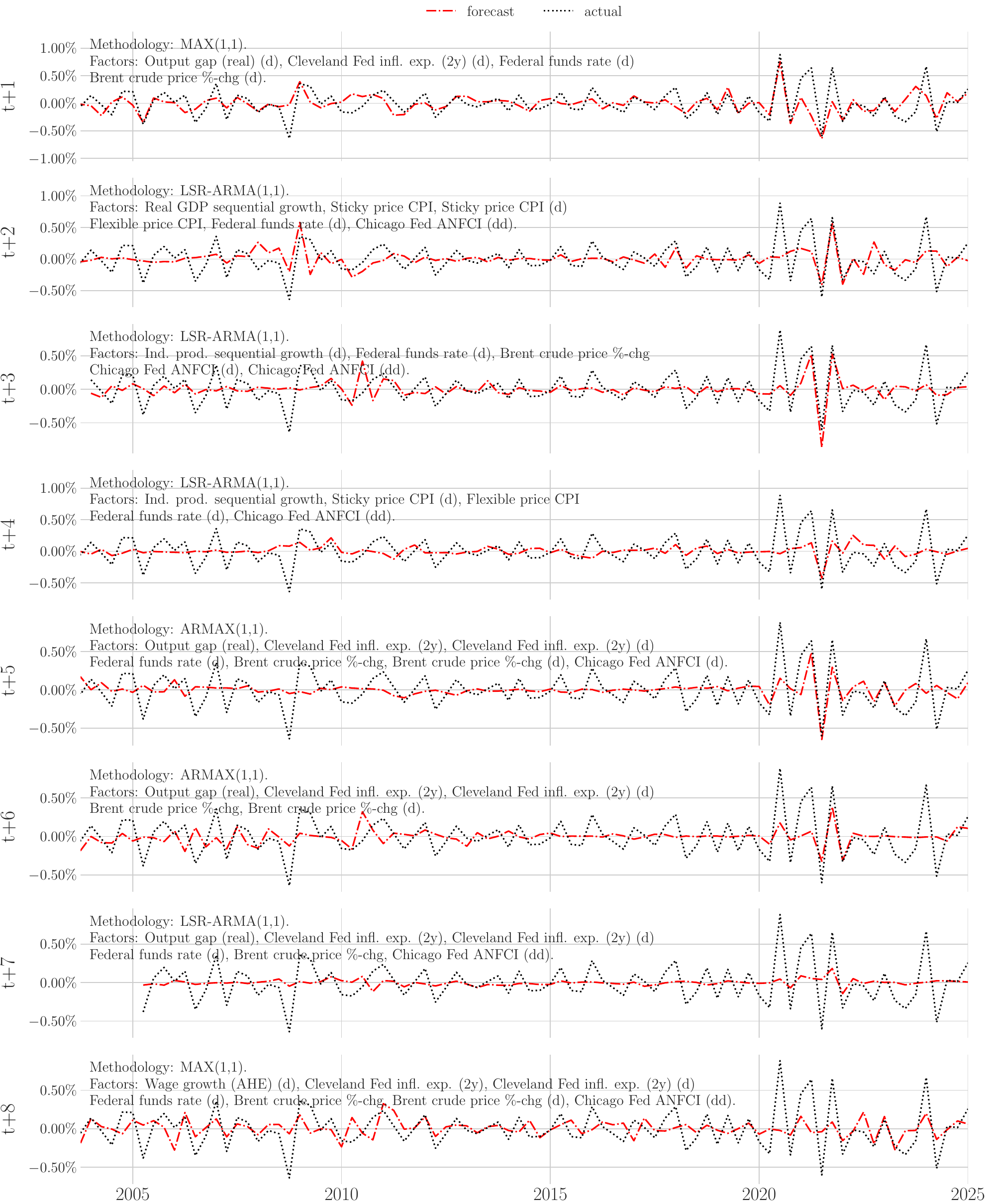}
    \label{fig-best_models_by_horizon_core_ma}
\end{figure*}

\begin{figure*}[ht]
    \caption{Average change in MSPE from including a predictor in an LSR-ARMA(1,1) PC model for core PCE}
    \includegraphics[width=\textwidth]{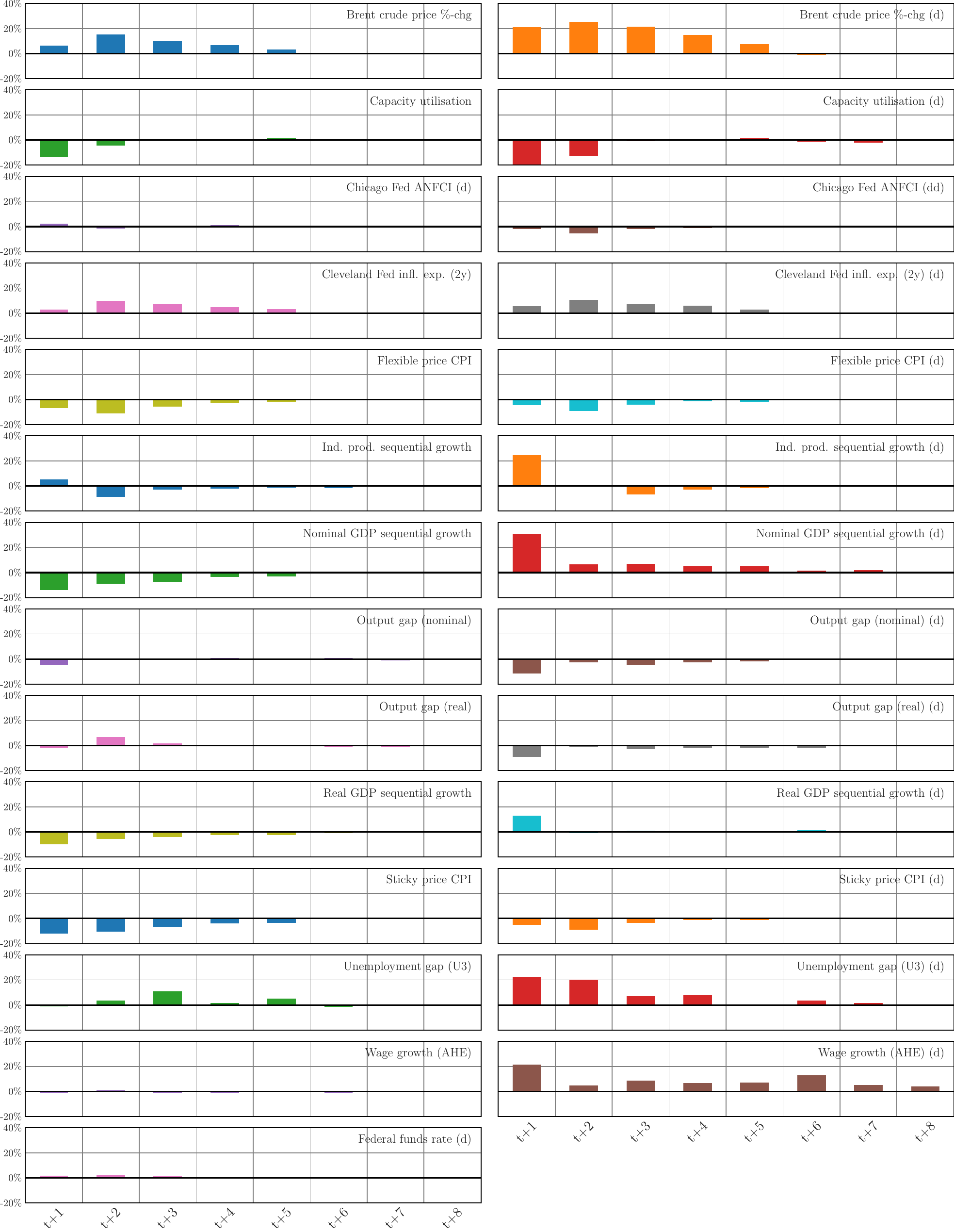}
    \label{fig-mspe_predictors_core_ma}
\end{figure*}

\FloatBarrier

\begin{figure*}[ht]
    \centering
    \caption{MSPE rankings, like-for-like factor specifications: Headline PCE}
    \includegraphics[width=0.85\textwidth]{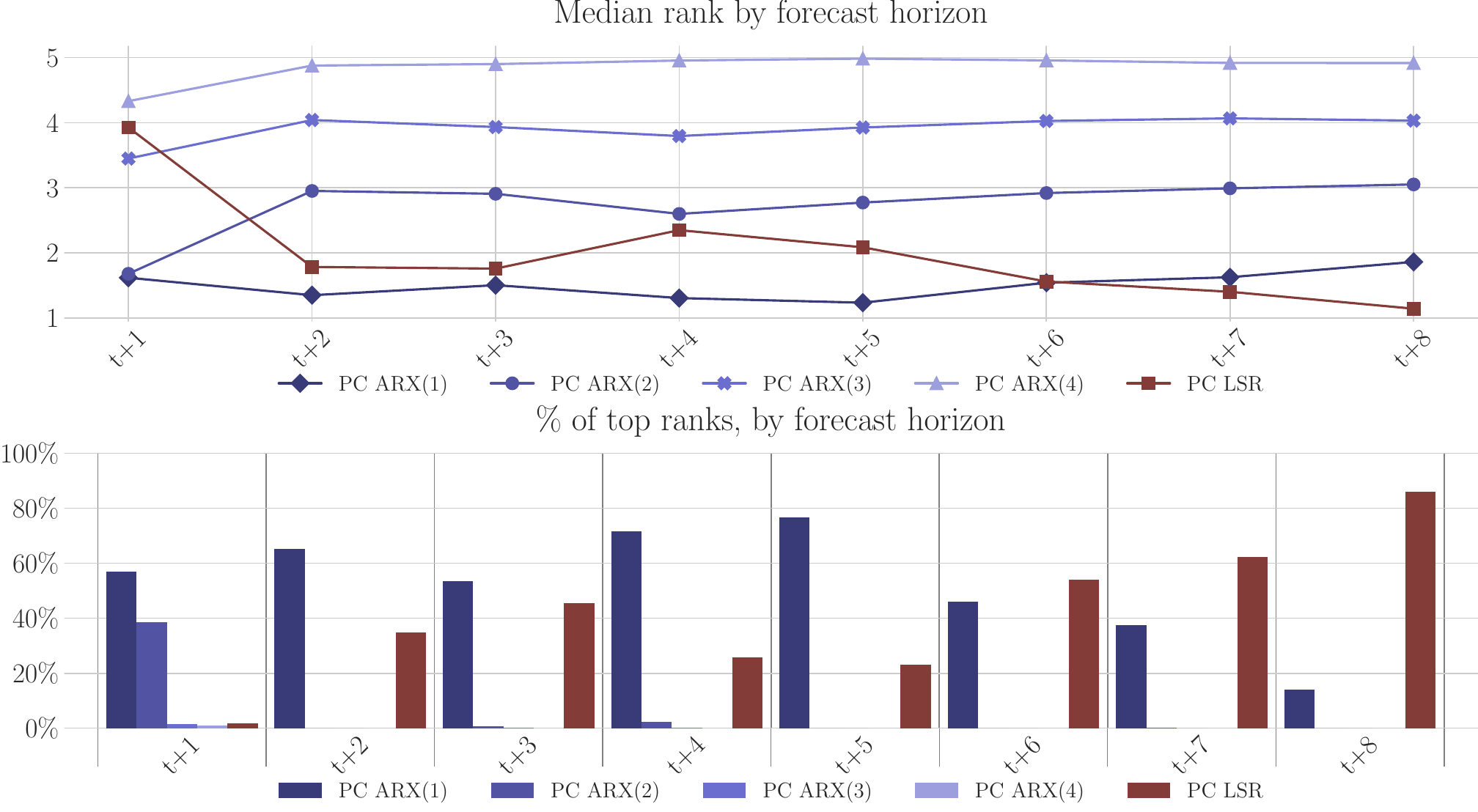}
    \label{fig-mspe_like_for_like_headline}
\end{figure*}

\begin{figure*}[h!]
    \centering
    \caption{MSPE rankings incl. univariate models and EWMA: Headline PCE}
    \includegraphics[width=0.85\textwidth]{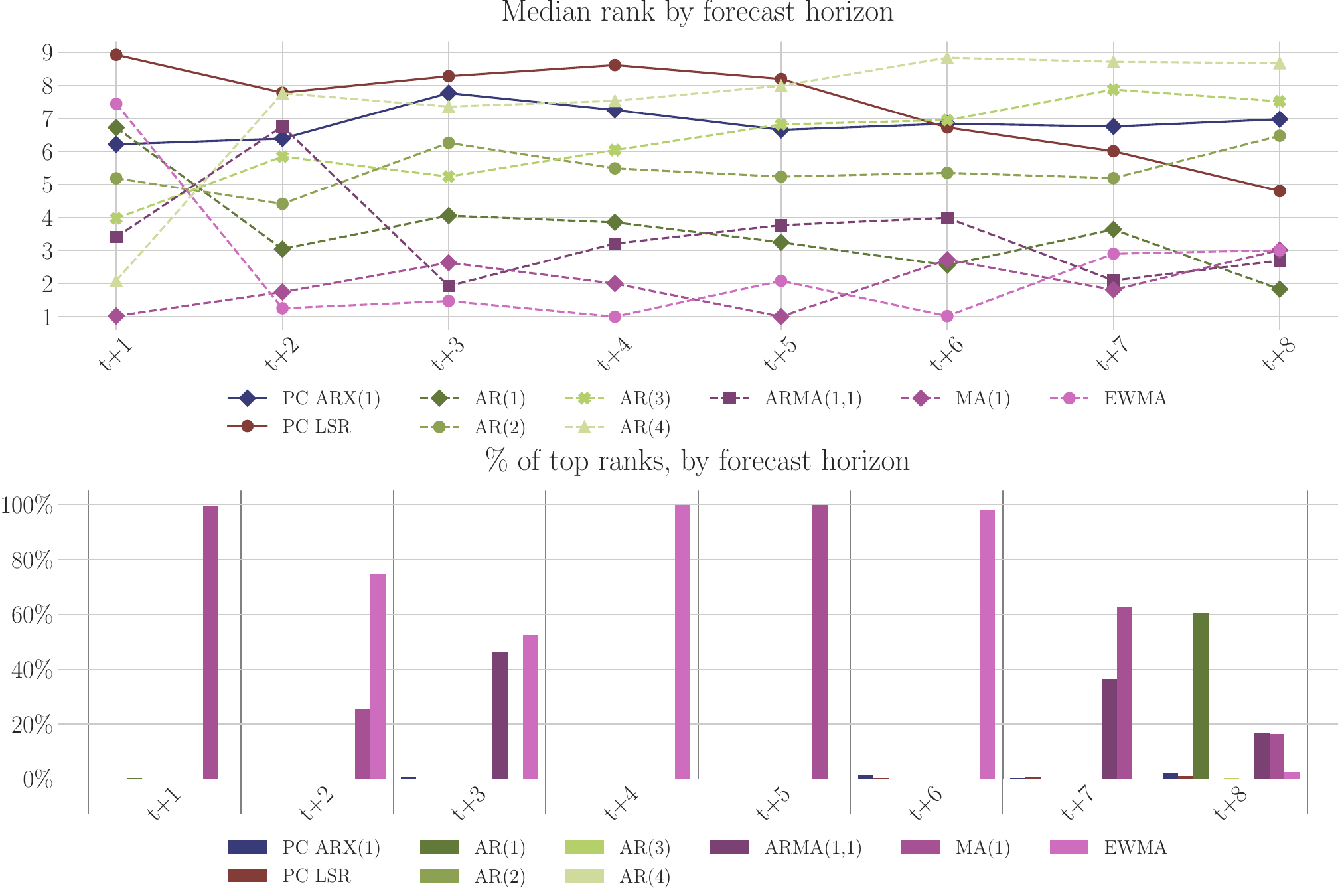}
    \label{fig-mspe_ranks_all_headline}
\end{figure*}

\FloatBarrier

\begin{figure*}[ht]
    \centering
    \caption{MSPE rankings after MA(1) adjustment of PC models: Headline PCE (1)}
    \includegraphics[width=0.85\textwidth]{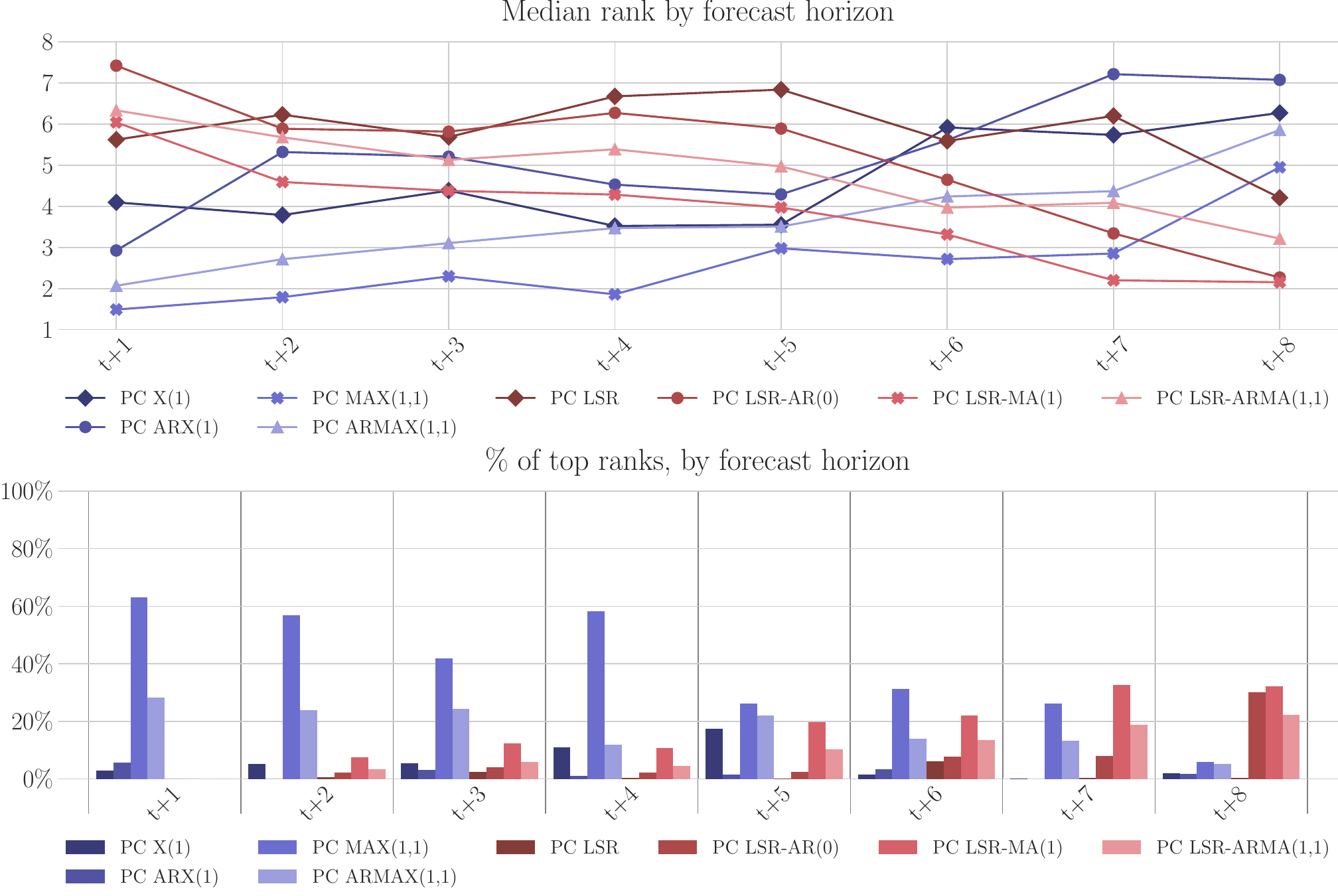}
    \label{fig-mspe_like_for_like_headline_ma}
\end{figure*}

\begin{figure*}[ht]
    \centering
    \caption{MSPE rankings after MA(1) adjustment of PC models: Headline PCE (2)}
    \includegraphics[width=0.85\textwidth]{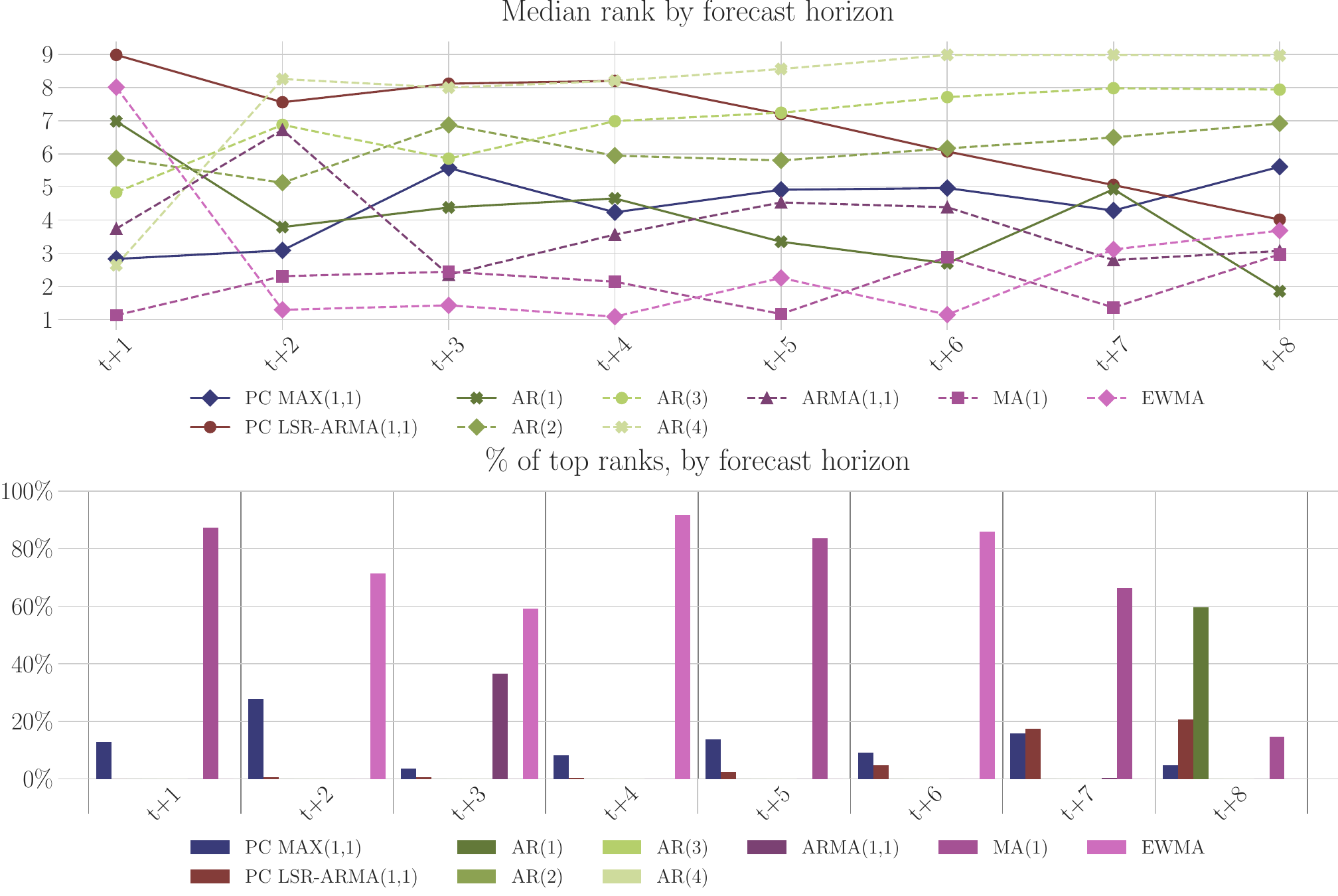}
    \label{fig-mspe_ranks_all_headline_ma}
\end{figure*}

\FloatBarrier

\bibliography{/home/daniil/Research/bibliography/phd1}

\begin{thebibliography}{35}
\expandafter\ifx\csname natexlab\endcsname\relax\def\natexlab#1{#1}\fi
\providecommand{\url}[1]{\texttt{#1}}
\providecommand{\href}[2]{#2}
\providecommand{\path}[1]{#1}
\providecommand{\DOIprefix}{doi:}
\providecommand{\ArXivprefix}{arXiv:}
\providecommand{\URLprefix}{URL: }
\providecommand{\Pubmedprefix}{pmid:}
\providecommand{\doi}[1]{\href{http://dx.doi.org/#1}{\path{#1}}}
\providecommand{\Pubmed}[1]{\href{pmid:#1}{\path{#1}}}
\providecommand{\bibinfo}[2]{#2}
\ifx\xfnm\relax \def\xfnm[#1]{\unskip,\space#1}\fi
\bibitem[{Albuquerque and Baumann(2017)}]{Albuquerque-2017}
\bibinfo{author}{Albuquerque, B.}, \bibinfo{author}{Baumann, U.},
  \bibinfo{year}{2017}.
\newblock \bibinfo{title}{Will us inflation awake from the dead? the role of
  slack and non-linearities in the phillips curve}.
\newblock \bibinfo{journal}{Journal of Policy Modeling} \bibinfo{volume}{39},
  \bibinfo{pages}{247--271}.
\newblock \URLprefix
  \url{https://www.sciencedirect.com/science/article/pii/S0161893817300066},
  \DOIprefix\doi{10.1016/j.jpolmod.2017.01.004}.
\bibitem[{Ang et~al.(2007)Ang, Bekaert and Wei}]{Ang-2007}
\bibinfo{author}{Ang, A.}, \bibinfo{author}{Bekaert, G.}, \bibinfo{author}{Wei,
  M.}, \bibinfo{year}{2007}.
\newblock \bibinfo{title}{Do macro variables, asset markets, or surveys
  forecast inflation better?}
\newblock \bibinfo{journal}{Journal of Monetary Economics}
  \bibinfo{volume}{54}, \bibinfo{pages}{1163--1212}.
\newblock \URLprefix
  \url{https://www.sciencedirect.com/science/article/pii/S0304393206002303},
  \DOIprefix\doi{10.1016/j.jmoneco.2006.04.006}.
\bibitem[{Ball(2000)}]{Ball-2000}
\bibinfo{author}{Ball, L.}, \bibinfo{year}{2000}.
\newblock \bibinfo{title}{Near-Rationality and Inflation in Two Monetary
  Regimes}.
\newblock \bibinfo{type}{Economics Working Paper Archive}
  \bibinfo{number}{435}. The Johns Hopkins University,Department of Economics.
\newblock \URLprefix \url{https://ideas.repec.org/p/jhu/papers/435.html}.
\bibitem[{Ball et~al.(1988)Ball, Mankiw, Romer, Akerlof, Rose, Yellen and
  Sims}]{Ball-1988}
\bibinfo{author}{Ball, L.}, \bibinfo{author}{Mankiw, N.G.},
  \bibinfo{author}{Romer, D.}, \bibinfo{author}{Akerlof, G.A.},
  \bibinfo{author}{Rose, A.}, \bibinfo{author}{Yellen, J.},
  \bibinfo{author}{Sims, C.A.}, \bibinfo{year}{1988}.
\newblock \bibinfo{title}{The new keynesian economics and the output-inflation
  trade-off}.
\newblock \bibinfo{journal}{Brookings Papers on Economic Activity}
  \bibinfo{volume}{1988}, \bibinfo{pages}{1--82}.
\newblock \URLprefix \url{http://www.jstor.org/stable/2534424}.
\bibitem[{Ball and Mazumder(2011)}]{Ball-2011}
\bibinfo{author}{Ball, L.M.}, \bibinfo{author}{Mazumder, S.},
  \bibinfo{year}{2011}.
\newblock \bibinfo{title}{Inflation Dynamics and the Great Recession}.
\newblock \bibinfo{type}{Working Paper} \bibinfo{number}{17044}. National
  Bureau of Economic Research.
\newblock \URLprefix \url{http://www.nber.org/papers/w17044},
  \DOIprefix\doi{10.3386/w17044}.
\bibitem[{Bargman(2025)}]{Bargman-clarx-2025-preprint}
\bibinfo{author}{Bargman, D.}, \bibinfo{year}{2025}.
\newblock \bibinfo{title}{Latent variable autoregression with exogenous
  inputs}.
\newblock \URLprefix \url{https://arxiv.org/abs/2506.04488},
  \href{http://arxiv.org/abs/2506.04488}{{\tt arXiv:2506.04488}}.
\bibitem[{Berriel et~al.(2016)Berriel, Medeiros and Sena}]{Berriel-2016}
\bibinfo{author}{Berriel, T.}, \bibinfo{author}{Medeiros, M.C.},
  \bibinfo{author}{Sena, M.J.}, \bibinfo{year}{2016}.
\newblock \bibinfo{title}{Instrument selection for estimation of a
  forward-looking phillips curve}.
\newblock \bibinfo{journal}{Economics Letters} \bibinfo{volume}{145},
  \bibinfo{pages}{123--125}.
\newblock \URLprefix
  \url{https://www.sciencedirect.com/science/article/pii/S0165176516301987},
  \DOIprefix\doi{10.1016/j.econlet.2016.05.032}.
\bibitem[{Bryan and Meyer(2010)}]{Bryan-2010}
\bibinfo{author}{Bryan, M.F.}, \bibinfo{author}{Meyer, B.},
  \bibinfo{year}{2010}.
\newblock \bibinfo{title}{Are some prices in the cpi more forward looking than
  others? we think so}.
\newblock \bibinfo{journal}{Economic Commentary, Federal Reserve Bank of
  Cleveland} \URLprefix
  \url{https://www.atlantafed.org/-/media/documents/research/inflationproject/stickyprice/sticky-price-cpi-supplemental-reading.pdf}.
\bibitem[{Campbell and Thompson(2007)}]{Campbell-2007}
\bibinfo{author}{Campbell, J.Y.}, \bibinfo{author}{Thompson, S.B.},
  \bibinfo{year}{2007}.
\newblock \bibinfo{title}{Predicting excess stock returns out of sample: Can
  anything beat the historical average?}
\newblock \bibinfo{journal}{The Review of Financial Studies}
  \bibinfo{volume}{21}, \bibinfo{pages}{1509--1531}.
\newblock \URLprefix \url{https://doi.org/10.1093/rfs/hhm055},
  \DOIprefix\doi{10.1093/rfs/hhm055},
  \href{http://arxiv.org/abs/https://academic.oup.com/rfs/article-pdf/21/4/1509/24453399/hhm055.pdf}{{\tt
  arXiv:https://academic.oup.com/rfs/article-pdf/21/4/1509/24453399/hhm055.pdf}}.
\bibitem[{Casarin et~al.(2025)Casarin, Peruzzi and Raggi}]{Casarin-2025}
\bibinfo{author}{Casarin, R.}, \bibinfo{author}{Peruzzi, A.},
  \bibinfo{author}{Raggi, D.}, \bibinfo{year}{2025}.
\newblock \bibinfo{title}{Multiple equilibria and the phillips curve: Do agents
  always underreact?}
\newblock \bibinfo{journal}{Ca' Foscari University of Venice, Department of
  Economics Research Paper Series} \URLprefix
  \url{https://ssrn.com/abstract=5368667},
  \DOIprefix\doi{10.2139/ssrn.5368667}.
\bibitem[{Chen et~al.(2014)Chen, Turnovsky and Zivot}]{Chen-2014}
\bibinfo{author}{Chen, Y.c.}, \bibinfo{author}{Turnovsky, S.J.},
  \bibinfo{author}{Zivot, E.}, \bibinfo{year}{2014}.
\newblock \bibinfo{title}{Forecasting inflation using commodity price
  aggregates}.
\newblock \bibinfo{journal}{Journal of Econometrics} \bibinfo{volume}{183},
  \bibinfo{pages}{117--134}.
\newblock \URLprefix
  \url{https://www.sciencedirect.com/science/article/pii/S0304407614001547},
  \DOIprefix\doi{10.1016/j.jeconom.2014.06.013}. \bibinfo{note}{internally
  Consistent Modeling, Aggregation, Inference and Policy}.
\bibitem[{Conti(2021)}]{Conti-2021}
\bibinfo{author}{Conti, A.M.}, \bibinfo{year}{2021}.
\newblock \bibinfo{title}{Resurrecting the phillips curve in low-inflation
  times}.
\newblock \bibinfo{journal}{Economic Modelling} \bibinfo{volume}{96},
  \bibinfo{pages}{172--195}.
\newblock \URLprefix
  \url{https://www.sciencedirect.com/science/article/pii/S026499932031261X},
  \DOIprefix\doi{10.1016/j.econmod.2020.11.019}.
\bibitem[{Eisenstat and Strachan(2016)}]{Eisenstat-2016}
\bibinfo{author}{Eisenstat, E.}, \bibinfo{author}{Strachan, R.W.},
  \bibinfo{year}{2016}.
\newblock \bibinfo{title}{Modelling inflation volatility}.
\newblock \bibinfo{journal}{Journal of Applied Econometrics}
  \bibinfo{volume}{31}, \bibinfo{pages}{805--820}.
\newblock \URLprefix
  \url{https://onlinelibrary.wiley.com/doi/abs/10.1002/jae.2469},
  \DOIprefix\doi{https://doi.org/10.1002/jae.2469},
  \href{http://arxiv.org/abs/https://onlinelibrary.wiley.com/doi/pdf/10.1002/jae.2469}{{\tt
  arXiv:https://onlinelibrary.wiley.com/doi/pdf/10.1002/jae.2469}}.
\bibitem[{Faust and Wright(2013)}]{Faust-2013}
\bibinfo{author}{Faust, J.}, \bibinfo{author}{Wright, J.H.},
  \bibinfo{year}{2013}.
\newblock \bibinfo{title}{Chapter 1 - forecasting inflation}, in:
  \bibinfo{editor}{Elliott, G.}, \bibinfo{editor}{Timmermann, A.} (Eds.),
  \bibinfo{booktitle}{Handbook of Economic Forecasting}.
  \bibinfo{publisher}{Elsevier}. volume~\bibinfo{volume}{2} of
  \textit{\bibinfo{series}{Handbook of Economic Forecasting}}, pp.
  \bibinfo{pages}{2--56}.
\newblock \URLprefix
  \url{https://www.sciencedirect.com/science/article/pii/B9780444536839000013},
  \DOIprefix\doi{10.1016/B978-0-444-53683-9.00001-3}.
\bibitem[{Friedman(1960)}]{Friedman-1960-book}
\bibinfo{author}{Friedman, M.}, \bibinfo{year}{1960}.
\newblock \bibinfo{title}{A program for monetary stability}.
\newblock \bibinfo{number}{3}, \bibinfo{publisher}{Ravenio Books}.
\bibitem[{Gordon(1977)}]{Gordon-1977a}
\bibinfo{author}{Gordon, R.J.}, \bibinfo{year}{1977}.
\newblock \bibinfo{title}{The theory of domestic inflation}.
\newblock \bibinfo{journal}{The American Economic Review} \bibinfo{volume}{67},
  \bibinfo{pages}{128--134}.
\newblock \URLprefix \url{http://www.jstor.org/stable/1815895}.
\bibitem[{Gordon(2011)}]{Gordon-2011}
\bibinfo{author}{Gordon, R.J.}, \bibinfo{year}{2011}.
\newblock \bibinfo{title}{The history of the phillips curve: Consensus and
  bifurcation}.
\newblock \bibinfo{journal}{Economica} \bibinfo{volume}{78},
  \bibinfo{pages}{10--50}.
\newblock \URLprefix
  \url{https://onlinelibrary.wiley.com/doi/abs/10.1111/j.1468-0335.2009.00815.x},
  \DOIprefix\doi{10.1111/j.1468-0335.2009.00815.x},
  \href{http://arxiv.org/abs/https://onlinelibrary.wiley.com/doi/pdf/10.1111/j.1468-0335.2009.00815.x}{{\tt
  arXiv:https://onlinelibrary.wiley.com/doi/pdf/10.1111/j.1468-0335.2009.00815.x}}.
\bibitem[{Hammond and Gordon(2012)}]{Hammond-2012}
\bibinfo{author}{Hammond, G.}, \bibinfo{author}{Gordon, R.J.},
  \bibinfo{year}{2012}.
\newblock \bibinfo{title}{State of the art of inflation targeting}.
\newblock Number~\bibinfo{number}{29} in \bibinfo{series}{Handbooks},
  \bibinfo{publisher}{Centre for Central Banking Studies, Bank of England}.
\newblock \URLprefix \url{https://ideas.repec.org/b/ccb/hbooks/29.html}.
\bibitem[{Hannan and Rissanen(1982)}]{hannan-rissanen-1982}
\bibinfo{author}{Hannan, E.J.}, \bibinfo{author}{Rissanen, J.},
  \bibinfo{year}{1982}.
\newblock \bibinfo{title}{Recursive estimation of mixed autoregressive-moving
  average order}.
\newblock \bibinfo{journal}{Biometrika} \bibinfo{volume}{69},
  \bibinfo{pages}{81--94}.
\newblock \URLprefix \url{http://www.jstor.org/stable/2335856}.
\bibitem[{Hommes et~al.(2023)Hommes, Mavromatis, Özden and Zhu}]{Hommes-2023}
\bibinfo{author}{Hommes, C.}, \bibinfo{author}{Mavromatis, K.},
  \bibinfo{author}{Özden, T.}, \bibinfo{author}{Zhu, M.},
  \bibinfo{year}{2023}.
\newblock \bibinfo{title}{Behavioral learning equilibria in new keynesian
  models}.
\newblock \bibinfo{journal}{Quantitative Economics} \bibinfo{volume}{14},
  \bibinfo{pages}{1401--1445}.
\newblock \URLprefix
  \url{https://onlinelibrary.wiley.com/doi/abs/10.3982/QE1533},
  \DOIprefix\doi{https://doi.org/10.3982/QE1533},
  \href{http://arxiv.org/abs/https://onlinelibrary.wiley.com/doi/pdf/10.3982/QE1533}{{\tt
  arXiv:https://onlinelibrary.wiley.com/doi/pdf/10.3982/QE1533}}.
\bibitem[{Hommes and Sorger(1998)}]{Hommes-1998}
\bibinfo{author}{Hommes, C.}, \bibinfo{author}{Sorger, G.},
  \bibinfo{year}{1998}.
\newblock \bibinfo{title}{Consistent expectations equilibria}.
\newblock \bibinfo{journal}{Macroeconomic Dynamics} \bibinfo{volume}{2},
  \bibinfo{pages}{287–321}.
\newblock \DOIprefix\doi{10.1017/S1365100598008013}.
\bibitem[{Hornstein(2024)}]{Hornstein-2024}
\bibinfo{author}{Hornstein, A.}, \bibinfo{year}{2024}.
\newblock \bibinfo{title}{Residual seasonality in monthly core inflation}.
\newblock \bibinfo{journal}{Richmond Fed Economic Brief} \bibinfo{volume}{24}.
\newblock \URLprefix
  \url{https://www.richmondfed.org/publications/research/economic_brief/2024/eb_24-13}.
\bibitem[{Lansing(2009)}]{Lansing-2009}
\bibinfo{author}{Lansing, K.J.}, \bibinfo{year}{2009}.
\newblock \bibinfo{title}{Time-varying u.s. inflation dynamics and the new
  keynesian phillips curve}.
\newblock \bibinfo{journal}{Review of Economic Dynamics} \bibinfo{volume}{12},
  \bibinfo{pages}{304--326}.
\newblock \URLprefix
  \url{https://www.sciencedirect.com/science/article/pii/S109420250800032X},
  \DOIprefix\doi{https://doi.org/10.1016/j.red.2008.07.002}.
\bibitem[{Lansing(2019)}]{Lansing-2019}
\bibinfo{author}{Lansing, K.J.}, \bibinfo{year}{2019}.
\newblock \bibinfo{title}{Improving the phillips curve with an interaction
  variable}.
\newblock \bibinfo{journal}{FRBSF Economic Letter} \bibinfo{volume}{13}.
\bibitem[{Mutascu(2019)}]{Mutascu-2019}
\bibinfo{author}{Mutascu, M.}, \bibinfo{year}{2019}.
\newblock \bibinfo{title}{Phillips curve in us: New insights in time and
  frequency}.
\newblock \bibinfo{journal}{Research in Economics} \bibinfo{volume}{73},
  \bibinfo{pages}{85--96}.
\newblock \URLprefix
  \url{https://www.sciencedirect.com/science/article/pii/S1090944318304010},
  \DOIprefix\doi{10.1016/j.rie.2019.01.007}.
\bibitem[{Nelder and Mead(1965)}]{nelder-mead-1965}
\bibinfo{author}{Nelder, J.A.}, \bibinfo{author}{Mead, R.},
  \bibinfo{year}{1965}.
\newblock \bibinfo{title}{A simplex method for function minimization}.
\newblock \bibinfo{journal}{The Computer Journal} \bibinfo{volume}{7},
  \bibinfo{pages}{308--313}.
\newblock \URLprefix \url{https://doi.org/10.1093/comjnl/7.4.308},
  \DOIprefix\doi{10.1093/comjnl/7.4.308}.
\bibitem[{Peach et~al.(2011)Peach, Rich and Cororaton}]{Peach-2011}
\bibinfo{author}{Peach, R.W.}, \bibinfo{author}{Rich, R.W.},
  \bibinfo{author}{Cororaton, A.}, \bibinfo{year}{2011}.
\newblock \bibinfo{title}{How does slack influence inflation?}
\newblock \bibinfo{journal}{Current Issues in Economics and Finance}
  \bibinfo{volume}{17}.
\newblock \URLprefix \url{https://ssrn.com/abstract=1895526},
  \DOIprefix\doi{10.2139/ssrn.1895526}.
\bibitem[{Phillips(1958)}]{Phillips-1958}
\bibinfo{author}{Phillips, A.W.}, \bibinfo{year}{1958}.
\newblock \bibinfo{title}{The relation between unemployment and the rate of
  change of money wage rates in the united kingdom, 1861-1957}.
\newblock \bibinfo{journal}{Economica} \bibinfo{volume}{25},
  \bibinfo{pages}{283--299}.
\newblock \URLprefix \url{http://www.jstor.org/stable/2550759}.
\bibitem[{Salisu and Isah(2018)}]{Salisu-2018}
\bibinfo{author}{Salisu, A.A.}, \bibinfo{author}{Isah, K.O.},
  \bibinfo{year}{2018}.
\newblock \bibinfo{title}{Predicting us inflation: Evidence from a new
  approach}.
\newblock \bibinfo{journal}{Economic Modelling} \bibinfo{volume}{71},
  \bibinfo{pages}{134--158}.
\newblock \URLprefix
  \url{https://www.sciencedirect.com/science/article/pii/S0264999317313111},
  \DOIprefix\doi{10.1016/j.econmod.2017.12.008}.
\bibitem[{Samuelson and Solow(1960)}]{Samuelson-Solow-1960}
\bibinfo{author}{Samuelson, P.A.}, \bibinfo{author}{Solow, R.M.},
  \bibinfo{year}{1960}.
\newblock \bibinfo{title}{Analytical aspects of anti-inflation policy}.
\newblock \bibinfo{journal}{The American Economic Review} \bibinfo{volume}{50},
  \bibinfo{pages}{177--194}.
\newblock \URLprefix \url{http://www.jstor.org/stable/1815021}.
\bibitem[{Sims(2006)}]{Sims-2006}
\bibinfo{author}{Sims, C.A.}, \bibinfo{year}{2006}.
\newblock \bibinfo{title}{Rational inattention: Beyond the linear-quadratic
  case}.
\newblock \bibinfo{journal}{American Economic Review} \bibinfo{volume}{96},
  \bibinfo{pages}{158–163}.
\newblock \URLprefix
  \url{https://www.aeaweb.org/articles?id=10.1257/000282806777212431},
  \DOIprefix\doi{10.1257/000282806777212431}.
\bibitem[{Stock and Watson(1999)}]{Stock-1999}
\bibinfo{author}{Stock, J.H.}, \bibinfo{author}{Watson, M.W.},
  \bibinfo{year}{1999}.
\newblock \bibinfo{title}{Forecasting inflation}.
\newblock \bibinfo{journal}{Journal of Monetary Economics}
  \bibinfo{volume}{44}, \bibinfo{pages}{293--335}.
\newblock \URLprefix
  \url{https://www.sciencedirect.com/science/article/pii/S0304393299000276},
  \DOIprefix\doi{10.1016/S0304-3932(99)00027-6}.
\bibitem[{Stock and Watson(2007)}]{Stock-2007}
\bibinfo{author}{Stock, J.H.}, \bibinfo{author}{Watson, M.W.},
  \bibinfo{year}{2007}.
\newblock \bibinfo{title}{Why has u.s. inflation become harder to forecast?}
\newblock \bibinfo{journal}{Journal of Money, Credit and Banking}
  \bibinfo{volume}{39}, \bibinfo{pages}{3--33}.
\newblock \URLprefix
  \url{https://onlinelibrary.wiley.com/doi/abs/10.1111/j.1538-4616.2007.00014.x},
  \DOIprefix\doi{10.1111/j.1538-4616.2007.00014.x}.
\bibitem[{Woodford(2011)}]{Woodford-2011}
\bibinfo{author}{Woodford, M.}, \bibinfo{year}{2011}.
\newblock \bibinfo{title}{Interest and prices : foundations of a theory of
  monetary policy}.
\newblock \bibinfo{edition}{1} ed., \bibinfo{publisher}{Princeton University
  Press}, \bibinfo{address}{Princeton}.
\bibitem[{Wu and Xia(2016)}]{Wu-Xia-2016}
\bibinfo{author}{Wu, J.C.}, \bibinfo{author}{Xia, F.D.}, \bibinfo{year}{2016}.
\newblock \bibinfo{title}{Measuring the macroeconomic impact of monetary policy
  at the zero lower bound}.
\newblock \bibinfo{journal}{Journal of Money, Credit and Banking}
  \bibinfo{volume}{48}, \bibinfo{pages}{253--291}.
\newblock \URLprefix
  \url{https://onlinelibrary.wiley.com/doi/abs/10.1111/jmcb.12300},
  \DOIprefix\doi{10.1111/jmcb.12300}.

\end{thebibliography}
\end{document}